\documentclass[a4paper,11pt]{article}
\pdfoutput=1 

\usepackage{jheppub} 
\usepackage{hyperref}
\usepackage{slashed}
\usepackage{verbatim}
\usepackage{rotate}

\usepackage[normalem]{ulem}
\usepackage{cancel}

\usepackage{graphicx,color}

\makeatletter
\def\@fpheader{\relax}
\makeatother

\newcommand{\be}{\begin{equation}}
\newcommand{\ee}{\end{equation}}
\newcommand{\bi}{\begin{itemize}}
\newcommand{\ei}{\end{itemize}}
\newcommand{\bea}{\begin{eqnarray}}
\newcommand{\eea}{\end{eqnarray}}
\newcommand{\bracket}[2]{\bra{#1}\,#2\rangle} 
\newcommand{\bra}[1]{\langle\,#1\,|}          
\newcommand{\ket}[1]{|\,#1\,\rangle}          
\newcommand{\ud}{\mathrm{d}}

\newcommand{\LCm}{{\scriptscriptstyle -}} 
\newcommand{\LCp}{{\scriptscriptstyle +}}
\newcommand{\LCpm}{{\scriptscriptstyle \pm}}

\newcommand{\LCperp}{{\scriptscriptstyle \perp}}

\newcommand{\lint}[1]{\int\!\ud{\sf #1}}
\newcommand{\sfpi}{\rotatebox[origin=b]{-90}{$\mathbf\vDash$}}

\usepackage[T1]{fontenc} \usepackage[latin1]{inputenc}

\title{Radiation reaction from QED: lightfront perturbation theory in a plane wave background}

\author{Anton Ilderton}
\author{and Greger Torgrimsson}
\affiliation{Department of Applied Physics \\ Chalmers University of Technology, SE-41296 Gothenburg, Sweden}

\emailAdd{anton.ilderton@chalmers.se}
\emailAdd{greger.torgrimsson@chalmers.se}

\abstract{We derive dynamical, real time radiation reaction effects from lightfront QED. Combining the Hamiltonian formalism with a plane wave background field, the calculation is performed in the Furry picture for which the background is treated exactly while interactions between quantum fields are treated in perturbation theory as normal. We work to a fixed order in perturbation theory, but no other approximation is made. The literature contains many proposals for the correct classical equation describing a radiating particle; we take the classical limit of our results and identify which equations are consistent with QED.}

\begin{document} 
\maketitle
\flushbottom

\section{Introduction}\label{intro}
The motion of a particle is influenced not only by external forces but also by the particle's own emission of radiation. Momentum conservation implies that the particle recoils when it radiates, an effect called `radiation reaction' (RR).

A rough order-of-magnitude estimate, equating the electron rest energy to the work done by an electric field over the classical electron radius, suggests that RR becomes significant at electric field strengths of roughly~$\sim10^{20}$V/m (though see below). This is two orders of magnitude higher than the Sauter-Schwinger limit at which nonperturbative QED effects come into play~\cite{Sauter:1931zz,Schwinger:1951nm}. Despite this, and the implied difficulty in observation, understanding RR presents one of the oldest and most frequently revisited problems in electrodynamics. Let us recall why.

Beginning with the coupled classical equations of motion for a particle, electromagnetic fields and external forces, it is possible to integrate out the field variables and write down an equation for the orbit of a radiating particle. This is the well known Lorentz-Abraham-Dirac (`LAD') equation \cite{L,A,D}, which has two unusual features. The first is that a divergence arises in its derivation, but this can be removed by renormalising the particle mass. The second feature of LAD is that it is third order in derivatives and admits unphysical runaway solutions in which even a free particle can spontaneously accelerate to the speed of light. Due to this, a great deal of work has over the years gone into deriving the `correct' equation of motion for a classical, radiating particle which avoids the problems of LAD.

Runaways are nonperturbative in the electromagnetic coupling, being given by $1/e^2$ terms~\cite{Coleman}. The simplest way to avoid them is therefore, given a particular system, to solve LAD perturbatively. To any given order the solutions are free from runaways and pre-acceleration (see, though~\cite{Zhang}). To obtain an equation which describes such physical solutions, one can reduce the order of LAD from third to second by recursively substituting the equation into itself, which eliminates higher derivative terms. To obtain a tractable equation one must additionally truncate this expansion to some order in the coupling. The Landau-Lifshitz equation (LL)~\cite{LL-bok} is the first order truncation, and is probably the most commonly employed equation to describe classical RR. The results of~\cite{Spohn:1999uf} support LL as the effective equation describing physical, i.e.\ non-runaway solutions to first order in the coupling\footnote{This result is sometimes misquoted as being exact, but it is clearly stated in~\cite{Spohn:1999uf} that higher orders are dropped. Further, LAD and LL differ already at order $e^4$, see below.}. Many further proposals for classical equations exist, and it is fair to say that there is no absolute consensus in the literature over which should be used to describe classical physics~\cite{Ror-Comment,O-Comment,Mer-Fritz}.

Perturbative quantum electrodynamics (QED) is the best tested theory of fundamental physics~\cite{PDG}. Starting in QED, we should be able to derive RR effects, and then take the classical limit to provide some definite answers. At first glance it seems straightforward to extend the above description of RR to quantum field theory. One begins with a state describing an electron, evolves this in time through, e.g., an external field which excites the state and causes photon emission, and then one measures the momentum of the electron as a function of time. This should show effects due to the electron recoiling as it emits. Fundamentally, though, quantum field theory is a multi-particle theory; an accelerated electron can radiate photons which can produce electron-positron pairs, and therefore  particle number is not conserved. The quantum notion of `an electron's recoil' then becomes ambiguous since there is a varying number of electrons in the system.

A formal way around this problem is to restrict to the regime in which there is only a single electron. If one begins with a single electron state then, in perturbation theory (in particular, in the Furry picture, see below), there remains a single electron in the system to order $\alpha=e^2/(4\pi\hbar)$; zeroth order terms contain only acceleration effects due to external forces, while order $\alpha$ terms describe photon emission and self-energy, both of which are reminiscent of classical RR~\cite{Coleman,Ilderton:2013tb}. At order $\alpha^2$, two-photon emission contributes, but so does pair production~\cite{Hu:2010ye,Ilderton:2010wr,King:2013osa}. To extend the idea of quantum RR to higher orders, one can work in a parameter regime for which pair production has much lower probability than multi-photon emission, see~\cite{DiPiazza:2010mv, Neitz} and below.

In a previous paper, we explained in general which diagrams contribute to lowest order RR effects in the $S$-matrix of QED~\cite{Ilderton:2013tb}; these are indeed photon emission and the electron self energy. It is only the inclusive combination of the two which yields a physical, measurable, IR-finite observable. We also showed explicitly, for a certain class of background fields, that one recovers known asymptotic results for classical RR from QED, in the limit $\hbar\to 0$: asymptotically, and to lowest order, all the classical equations we will consider agree, and are consistent with Larmor's formula (for quantum corrections to which see~\cite{Higuchi:2009ms,Dinu:2013hsd}). 

The purpose of this paper is to derive dynamical, real time (i.e.\ non-asymptotic) RR effects from QED, which requires going beyond $S$-matrix elements and instead investigating the dynamics of states at finite time. This is most natural in a Hamiltonian formalism, which we combine here with lightfront field theory~\cite{Brodsky:1997de,Heinzl:2000ht}, in which quantisation is performed on null hyperplanes and $x^\LCp = x^0 + x^3$ is the time co-ordinate. Going from $S$-matrix elements to finite time dynamics brings its own challenges; renormalisation in the Hamiltonian formalism, which we encounter below, is made difficult due to a lack of explicit covariance, a problem compounded on the lightfront due to the appearance of nonlocal, momentum-dependent counterterms~\cite{Mustaki:1990im}. Thus the investigation presented here goes somewhat beyond our previous, $S$-matrix based, calculation~\cite{Ilderton:2013tb}.

We will not try to derive a general equation from QED, but will instead consider a particular choice of background field. The main reason for this is that we wish to present a calculation with the smallest possible number of approximations and assumptions. The topic of RR is old, the literature vast, and there are many different approaches in play, see~\cite{Spohn:2004ik,DiPiazza:2011tq,Hammond} for reviews and references. The approach we use is to compute averages of appropriate momentum/position operators, before taking their classical limits in order to compare with the predictions of various classical equations.

In our calculation, we will use the coupling expansion of QED to treat interactions between the quantum fields perturbatively, but, at each order of perturbation theory, all other parameters (including the coupling to the background) will be treated exactly, thus eliminating potential ambiguities. We work to first nontrivial order in the coupling, and explain the extension to higher orders. Our interest here is in theory, and in comparing classical and quantum results. A more detailed phenomenological investigation of the coupling expansion in the context of RR  order will be presented elsewhere~\cite{US3}. A closely related calculation was given in~\cite{Krivitsky:1991vt}. The advantage of that paper is that the background field is arbitrary, and hence one can (in principle) derive an equation from QED. This necessitates an entirely perturbative treatment, though, and leads to a somewhat involved calculation in which only one of the possible RR terms (in the classical equation) is recovered. Though our own calculation is rather more restricted, it has the advantage of making the physics clear, and we will be able to recover all classical RR terms for our choice of background field.

This background is a plane wave, or null field, of arbitrary temporal profile. As we will see, this background is particularly amenable to a lightfront treatment and allows us to easily distinguish between different classical equations. It is also the background lying behind much of the work on `strong field QED', which studies the use of intense laser light, or strong magnetic fields, for investigating physics both within the standard model and beyond. See ~\cite{Jaeckel:2010ni,Redondo:2010dp, DiPiazza:2011tq} for recent reviews of this active topic. The intense laser fields which are now, or soon will be, available present a huge potential for observing RR~\cite{DiPiazza:2011tq,Heinzl:2011ur,Bulanov:2012zv}; it has been estimated that RR effects are actually measurable at $10^{22}$--$10^{23}$W/cm$^2$~\cite{DiPiazza:2009zz,Harvey:2011dp,Bulanov-PRE}, corresponding to field strengths which are significantly lower than in the coarse estimate above, and which will be reached by next generation laser facilities.  

This paper is organised as follows. In Sect.~\ref{klass-sekt} we review various proposals for classical equations describing a radiating particle, and summarise their predictions for motion in a plane wave background. In Sect.~\ref{SEKT:LF} we perform the lightfront quantisation of QED in this background, and then construct and renormalise the electron momentum operator. The calculation of quantum and classical RR effects in the momentum and position of a particle, at finite time, is presented and discussed in Sect.~\ref{SEKT:RR}. Our conclusions are presented in Sect.~\ref{SEKT:SAMMANFATTNING}. We take $c=\epsilon_0=1$ throughout, but $\hbar\not=1$ unless otherwise stated.
\section{Classical radiation reaction}\label{klass-sekt}
We begin by collecting various proposals for equations describing a classical radiating particle in an external electromagnetic field $F_\text{ext}$. Our interest is not in the properties of these individual equations {\it per se}, but in comparing their predictions with the classical limit of QED. We therefore refer the reader to the original articles, cited below, for more details. Most of the proposed classical equations take the form
\be\label{eom}
	m\ddot{x}^\mu=eF_\text{ext}^{\mu\nu}\dot{x}_\nu+\frac{2}{3}\frac{e^2}{4\pi} R^\mu \;,
\ee
in which a dot denotes a derivative with respect to proper time $\tau$, and $R_\mu$ describes the radiation reaction force. We write  $f:=eF_\text{ext}/m$ from here on. Table~\ref{R-tabel} lists the forms of $R_\mu$ (in part using a compact notation in which Lorentz indices should be read from left to right) for the equations of Lorentz-Abraham-Dirac (LAD) \cite{L,A,D}, Landau-Lifshitz (LL) \cite{LL-bok}, Eliezer/Ford-O'Connell (EFO)~\cite{FO,Eliezer}, Mo-Papas (MP)~\cite{MP} and Herrera (H)~\cite{Herrera}. The equation originally proposed by Eliezer~\cite{Eliezer} was rederived by FO for a particle with structure, rather than a point particle~\cite{Ror-Comment,Baylis,Rib,O-Comment}. Some of the differences between LAD, MP and LL were investigated in~\cite{Rivera}. Our final classical equation is due to Sokolov~(S)~\cite{SOK-JETP}. In this case, the velocity $\dot{x}^\mu$ is {\it not} proportional to the momentum $q^\mu$, so the equation has a different form than (\ref{eom}). See Appendix~\ref{S-APP}. From here on we focus on the equations in Table~\ref{R-tabel}, and state corresponding results for S.
\begin{table}
\begin{center}
\begin{tabular}{|c||c|c|c|} \hline
 & $R_{\mu\nu}$ such that $R_\mu = R_{\mu\nu}\dot{x}^\nu$ & & $R_\mu$ \\ \hline\hline
LAD  & $\dddot{x}_\mu \dot{x}_\nu- \dot{x}_\mu \dddot{x}_\nu $ & $\to$ & $\dddot{x}+\ddot{x}^2\dot{x}$ \\[2pt]\hline
LL &$ \dot{f}_{\mu\nu} + (f^2_{\mu\sigma}\dot{x}^\sigma) \dot{x}_\nu - \dot{x}_\mu (f^2_{\nu\sigma}\dot{x}^\sigma)$  &  $\to $ & $\displaystyle \dot{f}\dot{x}+ff\dot{x}+(f\dot{x})^2\dot{x}$ \\[2pt]\hline
EFO &  $\frac{\ud}{\ud\tau}(f_{\mu\sigma}\dot{x}^\sigma)\dot{x}_\nu - \dot{x}_\mu \frac{\ud}{\ud\tau}(f_{\nu\sigma}\dot{x}^\sigma)  $ & $\to$ & $\displaystyle \dot{f}\dot{x}+f\ddot{x}+\ddot{x}f\dot{x}\dot{x}$ \\[2pt]\hline
MP &$ (f_{\mu\sigma}\ddot{x}^\sigma)\dot{x}_\nu - \dot{x}_\mu (f_{\nu\sigma}\ddot{x}^\sigma)$   &$\to$& $f\ddot{x}+\ddot{x}f\dot{x}\dot{x} $ \\[2pt]\hline
H &$ (f^2_{\mu\sigma}\dot{x}^\sigma) \dot{x}_\nu - \dot{x}_\mu (f^2_{\nu\sigma}\dot{x}^\sigma)$  &  $\to $ & $\displaystyle ff\dot{x}+(f\dot{x})^2\dot{x}$ \\[2pt]\hline
\end{tabular}
\end{center}
\caption{\label{R-tabel} The radiation reaction force for a particle in an external field $f:=eF_\text{ext}/m$. The left hand column shows the antisymmetric tensor form of the force. This antisymmetry implies the mass-shell condition $\dot{x}^2=1$, using which we simplify $R_\mu$ in the right hand column.}
\end{table}

We will solve the above equations using a classical analogue of perturbation theory in the Furry picture, a technique long used in strong-field-QED~\cite{Nikishov:1963,Nikishov:1964a}. (The quantum expansion is described in Section~\ref{SEKT:LF}.)  The idea is simple: we treat radiation reaction as a perturbation to the Lorentz force. We rewrite (\ref{eom}) as
\be\label{eom2}
	\ddot{x}^\mu=f^{\mu\nu}\dot{x}_\nu+\frac{2}{3}\frac{e^2}{4\pi m} R^\mu \;,
\ee
in which the coefficient of $R_\mu$ is the classical electron radius, $e^2/(4\pi m)\simeq 3$ fm. Noting that $e$ now appears only in the RR term, we expand the orbit in powers of $e^2$ (for simplicity of presentation; the appropriate dimensionless parameter is given below)
\be\label{x-utveckling}
	x(\tau)=x_0(\tau)+e^2  x_2(\tau)+e^4 x_4(\tau) +\ldots \;, 
\ee
in which subscripts indicate the power of $e$. We plug this into (\ref{eom}) and solve the equation in powers of $e^2$. To zeroth order, the equations of motion (\ref{eom2}) are
\be\label{Lorentz}
	\ddot{x}_0= f(x_0)\dot{x}_0\;,
\ee
which is the Lorentz force equation for a particle, orbit $x_0$, moving in a background field $F^\text{ext}_{\mu\nu}$. (This is the exact Lorentz force equation, not that equation expanded in the coupling.) Higher order terms $x_n$ in the expansion (\ref{x-utveckling}) correspond to modifications of the Lorentz orbit due to RR, i.e.\ due to the electron's recoil~\cite{Ilderton:2013tb}. While this expansion allows for a clear separation of Lorentz force and recoil effects, it is most useful when one can solve the Lorentz force equation exactly, as this allows higher order terms to be calculated analytically as well. This will be the case for our background field, to which we now turn\footnote{In our chosen background, exact solutions are actually available for LL~\cite{Exact} and S~\cite{SOK-JETP}. Since we are interested in comparing to perturbative QED, we need only the perturbative expansions of these results.}.

\subsection{Classical radiation reaction in null fields}
\begin{figure}[t]
\centering\includegraphics[width=0.5\textwidth]{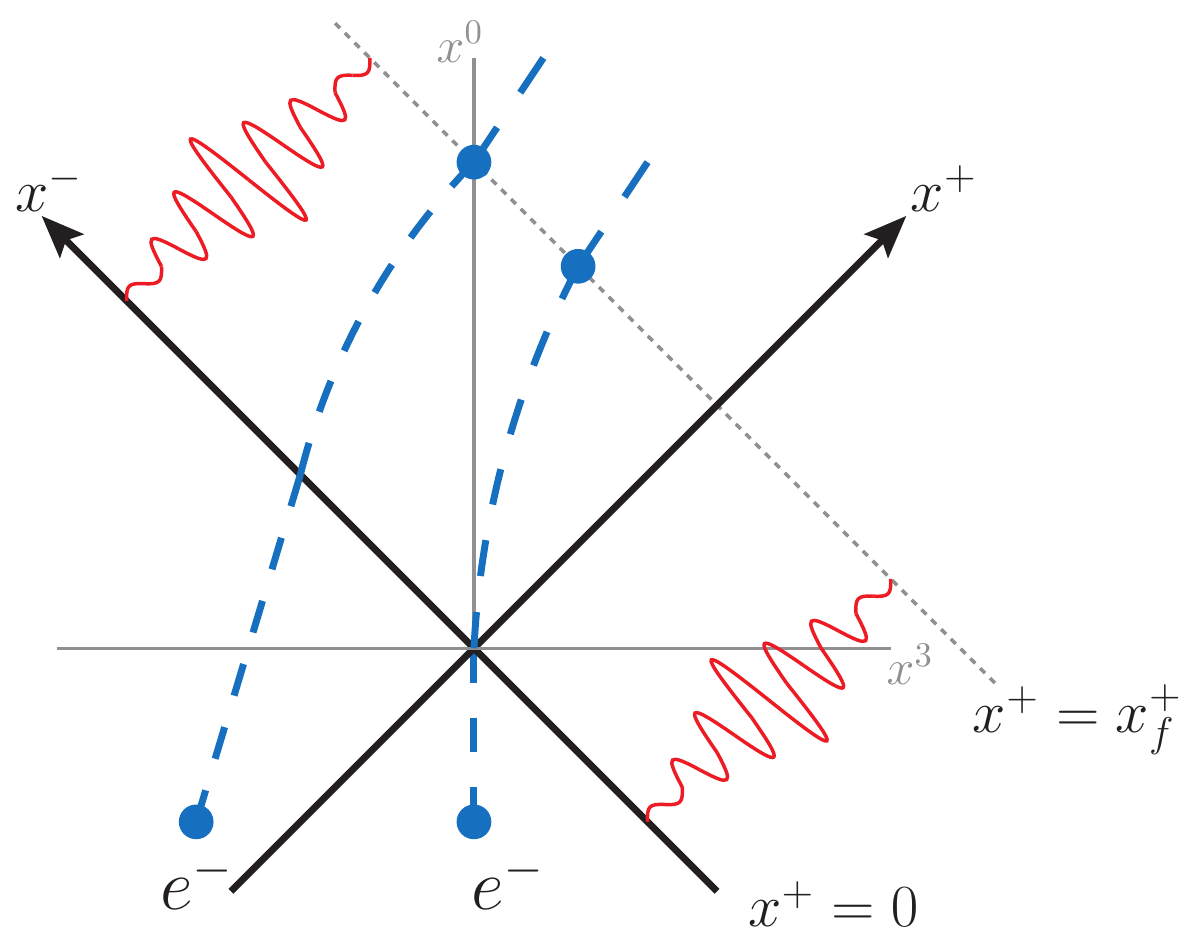}
\caption{\label{rumtid} A pulsed plane wave has finite extent in $x^\LCp$, but infinite extent in the remaining directions. All particles, independent of their momentum, enter and leave such a pulse at the same lightfront times, here $x^\LCp=0$ and $x^\LCp = x^\LCp_f$, respectively. There is no such symmetry in the usual time parameter $x^0$.}
\end{figure}

Our chosen background is a plane wave. This is constructed from a lightlike (null) wavevector $k_\mu$ and a transverse polarisation vector $a'_\mu$, so $ka'=0$. For the null vector we have $k_\mu=\omega n_\mu$, in which $\omega$ is an inverse-length scale, for example a central frequency, and we choose coordinates such that $nx = x^0 + x^3 = x^\LCp$, which is lightfront time~\cite{Brodsky:1997de,Heinzl:2000ht}. The field strength depends only on dimensionless invariant phase $\phi:=kx = \omega x^\LCp$ as
\be\label{F}
	eF^\text{ext}_{\mu\nu}(\phi)=k_\mu a'_\nu(\phi)-a'_\mu(\phi)k_\mu \;.
\ee
The polarisation vector and electric field are related by $a'_\LCperp(\phi) := e E_\LCperp(\phi) /\omega$. Throughout, a dash is a derivative with respect to $\phi$.  We refer to both $\phi$ and $x^\LCp$ as lightfront time. We consider pulses, so that $E_\LCperp(\phi)$ is either nonzero only in a finite $\phi$-range, or vanishes asymptotically, but is otherwise arbitrary. The field's spacetime structure is sketched in Fig.~\ref{rumtid}. All massive particles enter and leave a plane wave at the same lightfront time. Once they leave the pulse, they can never return to it. Further, taking $\phi$ as the time coordinate, spacetime acquires a `band structure', being cleanly separated into regions `before', `during' and `after' the pulse. No such separation is possible when using $x^0$ as the time coordinate, which manifests in properties of both the classical and quantum theories, see below. We now proceed to solve the classical equations of motion (\ref{eom2}) in a plane wave.

\subsubsection*{Zeroth order: Lorentz force}
To zeroth order, we need to solve (\ref{Lorentz}), the Lorentz force equation in a plane wave. The solution is well known, and follows from first observing that $k\ddot{x}_0 = 0$, since $kF_\text{ext}=0$, implying that $k\dot x_0$ is conserved. So, if the particle has momentum $p_\mu$ when it enters the pulse then, with $\phi_0 = kx_0$,
\be\label{phi0}
	\phi_0(\tau) = \frac{kp}{m} \tau \;.
\ee
The particle's proper time is, to zeroth order, proportional to lightfront time.  Using this, (\ref{Lorentz}) becomes a linear differential equation which can be solved immediately by exponentiation. The momentum $\pi:= m\dot{x}_0$ of a particle moving under the Lorentz force is
\be\label{pi-lorentz}
	\pi =p-a(\phi_0(\tau))+\frac{2a(\phi_0(\tau))p-a^2(\phi_0(\tau))}{2kp}k \;,
\ee
in which $a_\LCperp$ is the integral of the electric field strength\footnote{No gauge potential is used in the classical calculations.}, 
\be
	a_\LCperp(\phi) = \int\limits^\phi_{-\infty}\!\ud\varphi\ \frac{e}{\omega} E_\LCperp(\varphi) \;.
\ee
Here and below, the lower limit of the integral can be shifted to the point at which the field turns on; this is $-\infty$ for an asymptotically switched pulse, or can always be chosen to be $\phi=0$ for a compactly supported pulse, as in Fig.~\ref{rumtid}. To accommodate both options, we will frequently drop the lower limits on $\phi$ integrals, taking the relevant limit to be understood. From here on $p_\mu$ is always an initial momentum and $\pi_\mu$ always depends on $p_\mu$ as in (\ref{pi-lorentz}).

The solution (\ref{pi-lorentz}) may be integrated directly with respect to $\tau$ to find the position of the particle. Note that $a\equiv a(\phi_0(\tau))$, i.e.\ a function of $\phi_0$, which is a function of~$\tau$. This brings us to an important and slightly subtle point regarding the comparison between classical and quantum results, which we must discuss before calculating the first RR contribution to the orbit.

\subsubsection*{Reparameterisation}
The classical equations above are parameterised by proper time $\tau$. Solving them yields $x\equiv x(\tau)$, $\dot{x}(\tau)\equiv \partial_\tau x(\tau)$ and so on. In QED, and QFT in general, no reference is made to a particle worldline; since QFTs are multi-particle theories, a global time coordinate is used to parameterise state evolution~\cite{Dirac:1949cp}. Hence, any dynamical quantity obtained from QED will be parameterised by the time coordinate which defines the quantisation surfaces\footnote{It would though be interesting to reformulate our later calculations using worldline QED~\cite{Dunne:2005sx, Schubert:2001he}.}. In our case this will be $x^\LCp := x^0 + x^3$, which is natural given Fig.\ref{rumtid}. We therefore need to parameterise our classical solutions in terms of the {\it full} $x^\LCp$, or, to maintain covariance the full $\phi$ (and not $\tau$ or $\phi_0$). This can be done exactly for, at least, LL in a plane wave~\cite{Exact}, and in perturbation theory for all the considered equations.

To zeroth order, $\phi=\phi_0 = kp \tau/m$, so it is simple to eliminate $\tau$ in order to parameterise the momentum in terms of lightfront time $\phi$. We have
\be\label{pi-igen}
	m\dot{x}(\phi) =  p-a(\phi)+\frac{2pa(\phi)-a^2(\phi)}{2kp} + \mathcal{O}(e^2)  \equiv \pi(\phi) +\mathcal{O}(e^2)\;.
\ee
To compare the orbit with QED we can use (\ref{pi-igen}) to write, to zeroth order,
\be
	\frac{\ud x}{\ud \phi} 	=\frac{\dot x_0}{\dot \phi_0} + \mathcal{O}(e^2) = \frac{1}{kp} \pi(\phi)+ \mathcal{O}(e^2) \;,
\ee
following which a $\phi$-integration yields the orbit. At higher orders, inverting $\phi(\tau)$ to obtain $\tau(\phi)$ brings in more complicated functional dependencies since $\dot\phi$ is not constant when RR is accounted for~\cite{Exact}. (The implied symmetry breaking of null translation invariance provides a potential signal for measuring RR~\cite{Harvey:2011dp}.)

Even to zeroth order, there is no closed form parameterisation of the Lorentz orbit using instant-form time $x^0$. Mathematically, this is because $x^0(\tau)$ has no simple inverse, so we cannot explicitly eliminate $\tau$ in favour of $x^0$ to write $\tau = \tau(x^0)$. Physically, the reason is that the chosen background singles out $x^\LCp$ as a preferred direction\footnote{Note also that (\ref{pi-igen}) is the exact solution of the Lorentz force equation, and is the same as all previous (correct) expressions in the literature. This means that initial data is specified at a given initial $x^\LCp$ (which may be in the infinie past), not a given $x^0$. The analogous approach in the quantum theory is to use lightfront quantisation. We will do so in Section~\ref{SEKT:LF}.}, not $x^0$, as in Fig.~\ref{rumtid}.

\subsubsection*{First order: radiation reaction}
We are now ready to solve for $x_2$, the first correction to the Lorentz orbit due to RR. Inserting (\ref{x-utveckling}) into (\ref{eom2}), the equation to be solved is
\be\label{first-order-eq}
	\ddot{x}_2=f(\phi_0)\dot{x}_2+\phi_2 f'(\phi_0)\dot{x}_0+\frac{2}{3}\frac{1}{4\pi m}R(x_0) \;.
\ee
One may proceed as before: solve for $\phi_2$ by taking the inner product with $k_\mu$, following which (\ref{first-order-eq}) becomes a standard differential equation which can be solved for $\dot{x}_2^\mu$. Alternatively, one can begin with the $\{x^\LCm,x^\LCperp\}$ components (which are simple), and integrate directly with respect to $\tau$ or $\phi_0$, and then change variables to $\phi$; the final component, $x^\LCp$ then follows from the mass-shell condition. Either way, the result is that the momentum of the radiating particle is, to order $e^2$ and as a function of $\phi$,
\be\label{first-order}
	m\dot{x}(\phi)=\pi(\phi)+\frac{2}{3}\frac{e^2}{4\pi}\frac{m}{kp}\int\limits^\phi\!\ud\varphi\ \bigg(R(x_0(\varphi))-\frac{\pi(\phi)R(x_0(\varphi))}{kp}k\bigg) + \mathcal{O}(e^4) \;.
\ee
Either of the methods above can be extended directly to higher orders. It remains to insert the RR forces from Table~(\ref{R-tabel}) into  (\ref{first-order}). We summarise these results below. 

\subsection{Classical predictions}
Define the dimensionless parameter $\Delta$ by
\be
	\Delta:= \frac{2}{3}\frac{e^2}{4\pi}\frac{kp}{m^2} \;.
\ee
Inserting the expressions from Table~(\ref{R-tabel}) into (\ref{first-order}), one finds the momentum $q$ (equal to $m\dot{x}$ in all cases except S) to order $e^2$,
\be\label{LAD-P}\begin{split}
 \left\{\begin{array}{c} \text{LAD}  \\ \text{LL}  \\ \text{EFO} \end{array}\right\} \implies q(\phi) &=\pi(\phi)+\Delta \pi'(\phi)+\frac{\Delta}{m^2}\int\limits^\phi\!\ud\varphi\ a'^2(\varphi)\left(\pi(\varphi)-\frac{\pi(\varphi)\pi(\phi)}{kp}k\right)  \;, \\
	\left\{\begin{array}{c}  \text{MP} \\ \text{H} \\ \text{S}  \end{array}\right\} \implies q(\phi) &= q(\phi)_\text{LAD}-\Delta\pi'(\phi) \;. 
\end{split}
\ee
For the position (which follows from integration except in S, where $q\not= m\dot{x}$) we find
\begin{align}
\nonumber \left\{\begin{array}{c} \text{LAD}  \\ \text{LL}  \\ \text{EFO} \\ \text{S}\end{array}\right\} \implies kp\frac{\ud x(\phi)}{\ud\phi} &=\pi(\phi)+\Delta\pi'(\phi)+\frac{\Delta}{m^2}\int\limits^\phi\!\ud\varphi\ a'^2(\varphi)\bigg(\pi(\varphi)-\pi(\phi)-\frac{\pi(\varphi)\pi(\phi)}{kp}k\bigg)\;, \\
\label{LAD-X}	\left\{\begin{array}{c}  \text{MP} \\ \text{H}\end{array}\right\} \implies kp\frac{\ud x(\phi)}{\ud\phi}&= kp\frac{\ud x(\phi)}{\ud \phi}_\text{LAD}-\Delta\pi'(\phi) \;. 
\end{align}
The classical solutions differ in their transverse and longitudinal components. LAD, LL and EFO give the same result to order $e^2$, whereas MP, H and S predict a different result, as they do not contain the derivative $\Delta\pi'$. In this context, we note that LL is usually obtained by performing a reduction of order on LAD; applying the same reduction to EFO also yields LL~\cite{Hammond} (and hence these three equations agree, {\it to lowest order}, for all backgrounds). Reducing MP one finds H. Further, reducing MP as an equation for momentum $q$, one finds S, at least in a plane wave background. 

All the equations predict the same final momentum at $\phi=\infty$, which is equal to the momentum when the particle leaves the pulse; writing $\hat\pi\equiv \pi(\infty)$ from here on\footnote{$\hat\pi\not=p$ in general~\cite{unipolar2,Dinu:2012tj}, though attention is usually restricted to pulse shapes which do not give vacuum acceleration, in which case $\hat\pi=p$. All our results hold irrespective of the pulse shape.}, this is
\be\label{nb}
	q(\infty) = \hat{\pi} +\frac{\Delta}{m^2}\int\limits\!\ud\varphi\ a'^2(\varphi)\left(\pi(\varphi)-\frac{\pi(\varphi)\hat\pi}{kp}k\right)  \;.
\ee
This can also be derived from Larmor's formula for the total emitted radiation. The equations also agree, at all times, on the behaviour of the momentum component $k\dot{x}$, which is conserved in the case of the Lorentz force but with lowest order RR is
\be\label{asympt}
	kq(\phi) = kp\bigg(1+\frac{\Delta}{m^2} \int\limits^\phi\!\ud\varphi\ a'^2(\varphi)\bigg) +\mathcal{O}(e^4) \;.
\ee
From this non-conservation of $k\dot{x}$, we can extract the typical scale of RR effects in a plane wave. If we scale out $\eta = \tfrac{eE_0}{m\omega}$ as a typical magnitude of the electric field, and imagine that the pulse is short, roughly one cycle, so that the remaining $\phi$ integral in (\ref{asympt}) gives a factor of $2\pi$, we find
\be
	\frac{kq}{kp} \sim 1 - 2\pi\Delta \eta^2  +\mathcal{O}(e^4) = 1-\frac{2}{3}\frac{e^2}{2}\frac{kp}{m^2} \eta^2 +\mathcal{O}(e^4) \;.
\ee
($\Delta\eta^2$ is an appropriate, dimensionless, expansion parameter for RR in a plane wave.) For a 10 PW laser system, we have $\eta\sim 100$ and optical frequency $\hbar\omega\sim 1.24$ eV. Assuming a head-on collision between the beam and particle with gamma factor $\gamma$, we find
\be\label{uppfattning1}
	\frac{2}{3}\frac{e^2}{2}\frac{kp}{m^2} \eta^2 \sim 10^{-3}\gamma \;,
\ee
suggesting that RR effects reach, for example, $10\%$ of Lorentz force effects at around $m\gamma=0.03$~GeV. At $m\gamma\sim 0.3$ GeV, (\ref{uppfattning1}) reaches unity, and higher order corrections must be taken into account (see e.g.\ \cite{Exact} for the exact solution of LL in a plane wave.) This suggests that RR may indeed be measurable at the 10 PW laser facilities currently under construction.  For more detailed estimates and predictions of RR effects we refer the reader to~\cite{DiPiazza:2009zz,Harvey:2011dp,Bulanov-PRE}. While higher gamma factors can reduce field strengths required for observing RR, they also take us toward the quantum regime. Hence it is time to consider quantum effects. 

We will compare (\ref{LAD-P}) and (\ref{LAD-X}) to the classical limit of QED, specifically to expectation values of the electron momentum and position operators in the limit $\hbar\to 0$. Calculating these expectation values is the focus of the remainder of this paper.

\section{Lightfront quantisation of scalar QED}\label{SEKT:LF}
%
Our aim is to derive classical RR from the finite time dynamics of quantum states. We therefore use the Hamiltonian formalism. The calculation we will perform mirrors that in the classical theory, where we began with an electron and solved its equations of motion. Here we begin with single electron state and solve the Schr\"odinger equation for its time evolution. Also as above, we treat the `Lorentz component' exactly. For related calculations using the Hamiltonian approach and expectation values, see ~\cite{Krivitsky:1991vt}, `in-in' calculations~\cite{Johnson:2001rv,Johnson:2000qd,Galley:2010es} and position shift calculations~\cite{Higuchi:2002qc}. A comparison with $S$-matrix approaches is given below, and position calculations in our background are presented in Sect.~\ref{SEKT:X}. The calculation of~\cite{Krivitsky:1991vt}, in particular, is close in spirit to our own. Arbitrary background fields are treated, but only perturbatively. While this calculation can be extended to lightfront quantisation, the current approach with a simple background, treated exactly, is enough to obtain the results we are interested in.

Consider first a particle in an external plane wave, without radiation. We saw above that while the corresponding Lorentz equation can be solved exactly, the solution has no {\it explicit} parameterisation in terms of $x^0$. Parameterisation in terms of $\phi=kx$, or lightfront time, is on the other hand natural and straightforward. This is reflected in the quantum theory of a particle in a plane wave: despite the basics of this apparently simple theory having long been known~\cite{Volkov}, it is only in recent years that progress has been made in the instant form canonical quantisation of the theory~\cite{BOCA,Lavelle:2013wx}, and even then only for a specific choice of plane wave.  The situation is very different if one quantises on the lightfront~\cite{Brodsky:1997de, Heinzl:2000ht}. Quantisation then proceeds in analogy with the free theory, and has recently been used to clarify long-standing ambiguities regarding the `effective electron mass' in a plane wave~\cite{Ilderton:2012qe}. Further, when QED interactions are added, lightfront quantisation proceeds as for ordinary QED on the lightfront, as first shown in~\cite{Neville:1971uc}.

We consider scalar QED (sQED) from here on, as spin effects will in any case drop out in the classical limit. In order to perform the calculation of interest we must set up the theory, regulate and renormalise. sQED comprises the gauge field $A_\mu$, describing the photon, and a complex scalar field $\Phi$, describing the scalar electron and positron. Including an additional background $A_\mu^\text{ext}$, we have the action
\be
	S = \int\!\ud^4x\  -\frac{1}{4}F_{\mu\nu}F^{\mu\nu} + (D_\mu\Phi)^\dagger D^\mu\Phi-\tfrac{m^2}{\hbar}\Phi^\dagger \Phi \;,
\ee
in which $D_\mu = \partial_\mu + i\tfrac{e}{\hbar}A_\mu + i\tfrac{e}{\hbar}A^\text{ext}_\mu$, $F_{\mu\nu} = \partial_\mu A_\nu - \partial_\nu A_\mu$ and the background obeys Maxwell's equations in vacuum, $\partial_\mu F^{\mu\nu}_\text{ext}=0$. 

We will quantise on null hyperplanes of constant $\phi=kx$, on which the background field depends. This allows us to retain some explicit covariance~\cite{Walhout:1994rw}. We can, though, always choose our co-ordinates such that, as above, $kx=\omega(x^0+x^3)=\omega x^\LCp$, the usual lightfront time direction. With this choice the remaining coordinates are `transverse' $x^\LCperp=\{x^1,x^2\}$ and `longitudinal', $x^\LCm=x^0-x^3$. These will be denoted with sans-serif fonts, so ${\sf x} = \{x^\LCperp, x^\LCm\}$. Corresponding momentum components are ${\sf p} = \{p_\LCperp, p_\LCm\}$ with $p_\LCpm = \tfrac{1}{2}(p_0\pm p_3)$. Integrals over these variables are written
\be
	\int\!\ud {\sf x} := \int \ud x^\LCm \ud^2 x^\LCperp\;, \qquad \int\!\ud{\sf p} := \int\limits_0^\infty\!\frac{\ud p_\LCm}{(2\pi)2p_\LCm} \int\! \frac{\ud p_\LCperp}{(2\pi)^2} 
\ee
The derivation of the lightfront Hamiltonian follows that for sQED without background, so we highlight only the important steps. For a pedagogical discussion see~\cite{Brodsky:1997de}. For simplicity we set $\hbar=1$ and reintroduce it only before taking the classical limit. As usual, we use lightfront gauge, $A^\LCp \equiv 2 A_\LCm = 0$, for both the dynamical and background field. In lightfront quantisation, only the transverse components of $A_\mu$ are dynamical fields; the longitudinal component $A_\LCp$ is determined by $A_\LCperp$ and the current $J_\LCm$ according to\footnote{Inverting $\partial_\LCm$ requires a prescription for dealing with zero modes~\cite{Brodsky:1997de,Heinzl:2000ht,Heinzl:2003jy}, see also below.}
\be\label{constraint}
	A_\LCp = \frac{\partial_\LCperp A_\LCperp}{2\partial_\LCm} - \frac{J_\LCm}{2\partial_\LCm^2} \;.
\ee
The dynamical fields are then also transverse to the {\it background's} propagation direction, just like the physical fields of the background itself. This is one reason for orientating our quantisation surfaces with the laser direction~\cite{Walhout:1994rw}. A convenient choice of lightfront-gauge potential for the background field is then $eA_\mu^\text{ext}(x) = a_\mu(\phi)$, which is easily verified to give the correct field strength (\ref{F}). The usefulness of this choice will soon be clear. The Hamiltonian can now be written down,
\be\begin{split}\label{H-ALL}
	H &= \underset{H_0}{\underbrace{\frac{1}{2}\int\!\ud{\sf x}\ \frac{1}{2}A_j (i\partial_\LCperp)^2A_j +|\mathcal{D}_\LCperp\Phi|^2 +m^2|\Phi|^2}} \;,\\
	&+ \underset{V_1}{\underbrace{\frac{1}{2}\int\!\ud{\sf x}\ e j^\mu A_\mu}}\ \underset{V_2}{\underbrace{-e^2 A_\mu A^\mu |\Phi|^2 +\frac{e^2}{2}j_\LCm \frac{1}{(i\partial_-)^2}j_\LCm}} \;,
\end{split}
\ee
in which and from here on $A_\LCp \equiv \partial_\LCperp A_\LCperp / 2\partial_\LCm$~\cite{Brodsky:1997de}, the background-covariant derivative is $\mathcal{D}_\mu = \partial_\mu + i a_\mu$ and the background current is $j_\mu$,
\be
\begin{split}\label{current}
	j_\mu & = i\Phi^\dagger \mathcal{D}_\mu\Phi -i(\mathcal{D}_\mu\Phi)^\dagger \Phi  = \Phi^\dagger (i\overset{\leftrightarrow}{\partial}_\mu-2a_\mu)\Phi \;.
\end{split}
\ee
(Note that $J_\LCm=j_\LCm$.) Apart from the presence of $\mathcal{D}_\LCperp$ rather than $\partial_\LCperp$, (\ref{H-ALL}) is the ordinary Hamiltonian of lightfront sQED.  We will quantise in the Furry picture~\cite{Furry:1951zz}. This is a particular choice of interaction picture (which reduces to the usual one when the background vanishes) defined by the separation of the Hamiltonian into `free' and `interacting' parts.  The `free' theory is described by the first line of (\ref{H-ALL}), $H_0$. It comprises free photons, and the scalar field interacting with the plane wave. This theory can be solved exactly~\cite{Neville:1971uc}, like the classical Lorentz equation can be solved exactly. The second line of (\ref{H-ALL}) is the `interacting' part of the Hamiltonian, containing $V_1$ and $V_2$ which are respectively linear and quadratic in the coupling. These are the usual vertices of lightfront sQED but with the free matter current replaced by (\ref{current}); the three-point and four-point vertices, including instantaneous-scalar interactions, and finally the instantaneous-photon interaction. This `interacting' part of the Hamiltonian is treated in perturbation theory. In terms of position space Feynman diagrams, one has the usual vertices, the usual photon propagator, and a `dressed' fermion propagator describing the scalar's propagation within the background; if the background is turned off, $a_\mu=0$, the Furry picture reduces to the ordinary interaction picture of lightfront perturbation theory.

Operators evolve in the Furry picture under the action of the `free' Hamiltonian $H_0$. The transverse gauge field therefore has the usual mode expansion of lightfront field theory,
\be\label{A-expansion}
	A_j(x) = \int\!\ud {\sf l}\ a_j({\sf l}) e^{-il x} +a^\dagger_j({\sf l}) e^{il x} \;,
\ee
in which $l_\LCp = l_\LCperp^2/(4l_\LCm)$ as usual, so that $l^2 = 0$, on shell; all particles are on-shell in lightfront perturbation theory~\cite{Brodsky:1997de}. The fields obey the lightfront commutation relations~\cite{Heinzl:2000ht},
\be\label{LF-COMM}\begin{split}
	\big[A_i(x),A_j(y)\big]_{x^\LCp = y^\LCp} &= -\frac{i}{4}\delta_{ij} \varepsilon(x^\LCm - y^\LCm) \delta^{2}(x^\LCperp-y^\LCperp) \;, \\
	\big[a_i({\sf l}),a_j^\dagger({\sf l'})\big] &= 2 l_\LCm (2\pi)^3\delta^3({\sf l}-{\sf l'}) \delta_{ij} \;,
\end{split}
\ee
with $\varepsilon$ the sign function. Defining $a_\LCp({\sf l}) = l_\LCperp a_\LCperp({\sf l}) / 2l_\LCm$, the constrained field $A_\LCp$ can be written
\be
	A_\LCp(x)  = \int\!\ud {\sf l}\ a_\LCp({\sf l}) e^{-il x} +a^\dagger_\LCp({\sf l}) e^{il x}  \;,
\ee
and one can then write down a covariant expression for the mode commutators:
\be
	\big[a_\mu({\sf l}),a_\nu^\dagger({\sf l'})\big] = -2 l_\LCm (2\pi)^3\delta^3({\sf l}-{\sf l'}) \bigg(\eta_{\mu\nu} -\frac{k_\mu l_\nu + l_\mu k_\nu}{kl}\bigg) \equiv -2 l_\LCm (2\pi)^3\delta({\sf l}-{\sf l'})L_{\mu\nu}.
\ee
The scalar field operator, under the action of $H_0$, has the mode expansion
\be\label{phi-expansion}
	\Phi(x) =  \int\!\ud{\sf p}\ b({\sf p}) \varphi_{\sf p}(x) + d^\dagger({\sf p}) \varphi_{\sf -p}(x) \;, 
\ee
in which the mode functions $\varphi_{\sf p}$ are Volkov solutions~\cite{Volkov}, i.e.\ solutions to the Klein-Gordon equation in a background plane wave, 
\be
	\varphi_{\sf p}(x) = \exp\bigg[ -ipx -i\int\limits_0^{\phi}\! \frac{2 pa -a^2}{2kp}\bigg] \;,
\ee
with $p_\LCp=(p_\LCperp^2+m^2)/(4p_\LCm)$ so that $p^2=m^2$, on-shell. The mode operators are the (quantised) initial data on the hyperplane $\phi=0$, when the background turns on. $b^\dagger$ and $d^\dagger$ create on-shell scalar electrons and positrons with initial momentum $p_\mu$. One-particle states evolve under $H_0$ to carry the momenta $\pi_\mu$ of the classical theory, which is seen in the following property of the mode functions
\be
	\mathcal{D}_\mu\varphi_{\sf p}(x)=\pi_\mu(\phi)\varphi_{\sf p}(x) \;.
\ee
$\pi_\mu$ is of course also on-shell, $\pi_\LCp = (\pi_\LCperp^2+m^2)/(4p_\LCm)$, and $\pi_\LCm = p_\LCm$ is conserved under $H_0$--evolution, as was the case classically when considering only the Lorentz force. The scalar field obeys the commutation relations
\be\label{s-kom-1} \begin{split}
	\big[\Phi(x),\Phi^\dagger(y)\big]_{x^\LCp = y^\LCp} &= -\frac{i}{4}\varepsilon(x^\LCm - y^\LCm) \delta^{2}(x^\LCperp-y^\LCperp) \;, \\
	\big[b({\sf p}),b^\dagger({\sf q})\big] &= \big[d({\sf p}),d^\dagger({\sf q})\big] = 2 p_\LCm (2\pi)^3\delta^3({\sf p}-{\sf q}) \;.
\end{split}
\ee
We now turn to the states. These evolve in the Furry picture according to
\be
	i\frac{\partial}{\partial x^\LCp} \ket{\psi;x^\LCp}_F = H_F(x^\LCp)\ket{\psi;x^+}_F \;,
\ee
in which $H_F$ is the Furry analogue of the `interacting Hamiltonian in the interaction picture',
\be\label{HF-def}
	H_F = \mathcal{T^*_\LCp} e^{i\int\limits^{x^\LCp} H_0 } (V_1+V_2) \mathcal{T_\LCp} e^{-i\int\limits^{x^\LCp} H_0 }= (V_1+V_2) \big|_{A = (\ref{A-expansion}),\ \Phi = (\ref{phi-expansion})}
\ee
i.e.\ one inserts the `free field' mode expansions (\ref{A-expansion}) and (\ref{phi-expansion}) into the second line of (\ref{H-ALL}). We normal order throughout. Our interest lies in the case that the initial state describes a single electron of momentum $p_\mu$ in the far past, outside the pulse. In lightfront quantisation, the initial state is specified by the three components ${\sf p}=\{p_\LCperp,p_\LCm\}$ of the momentum $p_\mu$, rather than the three vector components ${\bf p}$, but the mass-shell condition $p^2=m^2$ shows that these are equivalent. The state is
\be\label{tillstaand-noll}
	\ket{\text{in}}=\lint{p}\ g({\sf p}) b^\dagger({\sf p})\ket{0} \;, \qquad 	\int\!\ud{\sf p}\ |g({\sf p})|^2 = 1 \;,
\ee
in which $\ket{0}$ is the lightfront vacuum annihilated by the $a({\sf l})$, $b({\sf p})$ and $d({\sf p})$,  and~$g({\sf p})$ is a wavepacket, normalised as shown\footnote{The form of the wavepacket must differ in the transverse and longitudinal directions, due to (technically) the lightfront momentum measure and (physically) the positivity of $p_\LCm$. We will return to this in Sect.~\ref{SEKT:X}.} such that $\bracket{\text{in}}{\text{in}}=1$, and strongly peaked around the momentum components $p_\LCperp$ and $p_\LCm$.  Our state at later lightfront time $x^\LCp$ is 
\be\label{tillstaand-phi}
	\ket{\psi;x^\LCp} = \mathcal{T}_\LCp \exp\bigg[-i\int\limits_{-\infty}^{x^\LCp}\!\ud s\ H_F(s)\bigg]\ket{\text{in}} \;,
\ee
in which we make the usual assumption of asymptotic switching to evolve the initial state up to the fully interacting electron state before it enters the background field. 

This completes our quantisation of the theory. We now want to calculate the expectation values of the electron momentum operator $P^e_\mu$ in the evolved state (\ref{tillstaand-phi}),
\be\label{P-EXP-FRI}
	\langle P^{e}_\mu \rangle = \bra{\psi;x^\LCp} P^e_\mu \ket{\psi;x^\LCp} \;,
\ee
and then take the classical limit in order to compare with the classical results in Sect.~\ref{klass-sekt}. First we need to write down the momentum operator.

\subsection{The momentum operator}
The energy-momentum tensor in our theory is
\be\label{T-TOT}
	T_{\mu\nu} = -\frac{2}{\sqrt{-g}}\frac{\delta S}{\delta g_{\mu\nu}} = F_{\mu\sigma}{F^{\sigma}}_\nu + (D_\mu\Phi)^\dagger  D_\nu \Phi +(D_\nu\Phi)^\dagger  D_\mu \Phi  - g_{\mu\nu}\mathcal{L} \;,
\ee
and the total momentum of the system is~\cite{Brodsky:1997de,Heinzl:2000ht}
\be
	P_\mu = \int\!\ud{\sf x}\ T_{\LCm\mu} \;,
\ee
in which $P_\LCp$ is the Hamiltonian (\ref{H-ALL}). We are not interested in the total momentum, but in that of the electron. We must therefore separate $T_{\mu\nu}$ into a piece which describes the electron, and a piece which describes the emitted radiation. This is trivial in the free theory, and also asymptotically in which the theory again becomes free\footnote{Modulo IR problems~\cite{Kulish:1970ut,Horan:1999ba}, but IR divergences drop out of our expectation value, see~\cite{Ilderton:2013tb} and below.}. At finite time, though, it is not obvious how to make such a separation, since the physical electron is a composite of the matter field and a cloud of photons~\cite{Dirac:1955uv,Lavelle:1995ty,Bagan:1999jf,Bagan:1999jk}. A related classical problem is the subject of energy balance, the question of how one separates the electromagnetic fields into bound and radiated parts, see~\cite{Gal'tsov:2010cz, Gal'tsov:2004qz} for detailed discussions.

There are several conditions which will help us to identify the relevant part of $T_{-\mu}$. First, the decomposition must be gauge invariant. Second, it must give the correct free-field limit. These two constraints (a third follows) suggest the natural electron-photon split  $T_{\mu\nu} = T^e_{\mu\nu} + T^\gamma_{\mu\nu}$ in which~\cite{Schott:1975si}
\be\label{T-split}\begin{split}
	T^e_{\mu\nu} &=   (D_\mu\Phi)^\dagger  D_\nu \Phi +(D_\nu\Phi)^\dagger  D_\mu \Phi  -g_{\mu\nu}\big( |D\Phi |^2-m^2|\Phi|^2\big) \;, \\
	T^\gamma_{\mu\nu}  &= F_{\mu\sigma} {F^{\sigma}}_\nu + g_{\mu\nu}\frac{1}{4} F_{\sigma\rho}F^{\sigma\rho}\;.
\end{split}\ee
These are the usual energy-momentum tensors for the electromagnetic field, and the minimally coupled scalar. This decomposition is gauge invariant and has the correct free-field limit, implying we should take
\be\label{P-split}
	P^e_\mu = \int\!\ud{\sf x}\ T^e_{\LCm\mu}  \;, \qquad	P^\gamma_\mu = \int\!\ud{\sf x}\ T^\gamma_{\LCm\mu} \;.
\ee
These operators are explicitly dependent on lightfront time due to the background field. To order $e^2$, antiparticles do not contribute to our calculation (a common feature of lightfront quantisation~\cite{Brodsky:1997de,Heinzl:2000ht}), and so we drop the positron modes $d({\sf p})$ in all our expressions. Similarly, some order $e^2$ terms in the operator drop out due to normal ordering. The {\it contributing} part of the electron momentum operator can then be be written as a sum of two terms, $P^e_\mu = P^{(0)}_\mu+P^{(1)}_\mu$, the first (second) being order zero (one) in the coupling:
\be\begin{split}\label{P-split-3}
	P^{(0)}_\mu(\phi) &= \int\!\ud{\sf p}\ \pi_\mu(\phi) b^\dagger({\sf p}) b({\sf p})  \;, \\
	P^{(1)}_\mu(\phi) &=  e \int\!\ud{\sf p}\ud{\sf l}\ b^\dagger({\sf p}-{\sf l}) b({\sf p})  \bigg( a^\dagger\pi(\phi) k_\mu - \frac{2kp-kl}{2}a_\mu^\dagger(l)\bigg)  \frac{e^{i\int^\phi\frac{l\pi}{kp-kl}}}{kp-kl} \;.
\end{split}
\ee
This is almost, but not quite, the operator we are looking for. Our third constraint is that $P^e_\mu$ should yield the correct sQED result: an electron with momentum $p_\mu$, unexposed to external forces, should always have momentum $p_\mu$. This means that we should find 
\be\label{renorm}
	\langle P^e_\mu\rangle \overset{?}{=} p_\mu \;.
\ee
Our first task is therefore to calculate $\langle P^e_\mu\rangle$ in ordinary sQED without a background field, and check that we obtain the correct answer (\ref{renorm}).

\subsection{On-shell renormalisation}\label{SEKT:RENORM}
%
Regularisation and renormalisation are necessary in the following calculations. We use dimensional regularisation to control UV-divergences. (For classical RR in different dimensions, see \cite{Galtsov:2001iv}.) Following \cite{Casher:1976ae}, the extra dimensions are placed into the transverse directions, so that
\be
	\int\!\ud^2 l_\LCperp\to \mu^{2\epsilon}\int\!\ud^nl_\LCperp \;,
\ee
where $n=2(1-\epsilon)$ and $\mu$ is the introduced mass scale. Transverse dim reg has the benefits of affecting neither the structure of the quantisation surfaces nor our chosen background; the plane wave is always homogeneous in the transverse directions, another reason for orientating our quantisation surfaces with the laser direction. See \cite{Heinzl:2006dw} for an application of transverse dim reg. Zero modes should be regulated using cutoffs, principle values or otherwise \cite{Brodsky:1997de,Heinzl:2000ht,Ligterink:1995wk}. We display the regulators only when necessary, but they are in place throughout. In the following calculation, dim reg is sufficient to take care of divergences. 

We set $a_\mu=0$ to return to ordinary lightfront perturbation theory. We then begin with an initial electron (\ref{tillstaand-noll}), and calculate the expectation value of its momentum at subsequent times, (\ref{P-EXP-FRI}) with $P^e$ as in (\ref{P-split-3}) but $a_\mu=0$. We will present the `in-background' version of this calculation in more detail later, and so we skip here to the final result. In the limit that the wave packet is strongly peaked around an initial momentum $p_\mu$, we find to order~$e^2$,
\be\begin{split}\label{renorm-mig0}
	\langle P^e_\mu \rangle = p_\mu +  e^2 &\int\!\ud{\sf l}\ \frac{kp-kl}{lp^2kp}\bigg(m^2-\frac{2kplp}{kl}\bigg)\bigg(l_\mu - \frac{lp}{kp-kl}k_\mu\bigg) \\
	&-\frac{2e^2}{kp}\int\ud{\sf l}\ \frac{2kp-kl}{2lp}\bigg(p_\mu - \frac{lp}{kl}k_\mu - \frac{kp}{kl}l_\mu\bigg) - \frac{1}{lp}\bigg(m^2-\frac{2kplp}{kl}\bigg)k_\mu \;.
\end{split}
\ee
The first line comes from $P^{(0)}$, the second line from $P^{(1)}$. We expect, from (\ref{renorm}), that the order $\alpha=e^2/(4\pi)$ terms here should vanish, and that the electron's momentum should remain unchanged. Evaluating the transverse integrals in (\ref{renorm-mig0}) using dim reg, we find that the order $\alpha$ terms are not zero, but divergent:
\be\label{renorm-mig}
	\langle P^e_\mu \rangle = p_\mu\bigg[ 1 - \frac{\alpha}{2\pi\epsilon}\bigg(\frac{\mu^2}{m^2}\bigg)^\epsilon+\ldots\bigg] + \frac{m^2}{kp}k_\mu \bigg[\frac{\alpha}{2\pi\epsilon}\bigg(\frac{\mu^2}{m^2}\bigg)^\epsilon+\ldots\bigg] \;,
\ee
where ellipses denote finite terms of order $\alpha$. (Dim reg is sufficient here; the longitudinal integrals do not contribute divergences.)  The physical result $p_\mu$ is multiplied by `one plus a divergent quantity', and we have a second divergence proportional to $k_\mu$. This second divergence is typical of lightfront quantisation, in that it depends on the momentum component $kp$~\cite{Mustaki:1990im,Brodsky:1997de,Heinzl:2000ht}. The result (\ref{renorm-mig}) is though covariant, and the second divergence depends only on the angular variables in $k_\mu$, as expected from~\cite{Walhout:1994rw}.

Removal of the divergences proceeds as follows. Recall that $S$-matrix elements are renormalised by adding counterterms to the Hamiltonian/Lagrangian. These counterterms renormalise the mass, charge and field normalisation, and are order $e^2$ or higher in sQED. However, we are not considering $S$-matrix elements, but expectation values of interacting operators at non-asymptotic times. Indeed, we find that, in our calculation, no order $e^2$ term in the interaction Hamiltonian contributes to the order $e^2$ expectation value. Counterterms in the interaction Hamiltonian therefore cannot remove the above divergences. Instead it is the composite operator $P^e_\mu$ itself which we need to renormalise~\cite{Collins}. We will nevertheless encounter some results familiar from lightfront perturbation theory.  See \cite{Dresti:2013kya} for closely related statements; that paper investigates expectation values and operator renormalisation in field theories where explicit time dependence is introduced not via a background field, as here, but by a time-dependent spacetime metric.

To proceed, we need a renormalisation condition. Given that a free electron of momentum $p_\mu$ should always have momentum $p_\mu$, none of the order $\alpha$ terms in (\ref{renorm-mig0}) and (\ref{renorm-mig}), divergent or finite, can be physical, and must be removed. We therefore take our renormalisation condition to be (\ref{renorm}). It seems natural to call this a Hamiltonian analogue of on-shell renormalisation. From here on we write down only divergent terms; finite terms are also removed by our renormalisation condition. Hence (\ref{renorm-mig}) becomes
\begin{equation*}
	\langle P^e_\mu \rangle = p_\mu\bigg[ 1 - \frac{\alpha}{2\pi\epsilon}+\ldots\bigg] + \frac{m^2}{kp}k_\mu \bigg[\frac{\alpha}{2\pi\epsilon}+\ldots\bigg] \;.
\end{equation*}
The first divergence is proportional to $p_\mu$, i.e.\ to the free theory result, and can therefore be removed by multiplicative operator renormalisation:
\be\label{P-renorm}
	P^e_\mu \to \bigg(1+\frac{\alpha}{2\pi\epsilon}\bigg) P^e_\mu \;.
\ee
In fact, the divergence comes entirely from the lowest order operator $P^{(0)}_\mu$, and hence only this part needs to be renormalised, but this is equivalent to (\ref{P-renorm}), to order $e^2$. We turn to the second divergence, proportional to $k_\mu$. This can be removed by mass renormalisation, and here we can make a connection to known results in lightfront renormalisation. Using transverse dim reg, the mass shift\footnote{Not to be confused with the `intensity dependent mass shift' in a plane wave, for which see \cite{Harvey:2011dp,Ilderton:2012qe}.} in lightfront quantisation is~\cite{Mustaki:1990im, Brodsky:1997de}
\be\label{dm}
	\delta m = \frac{\alpha}{2\pi\epsilon} \;.
\ee
Such a mass shift, of order $e^2$, in the Lagrangian could enter the order $e^2$ expectation value in only one way. As discussed above, it could not enter via the interaction Hamiltonian. Instead it can enter via the momentum operator which, recall, contains the term $k_\mu\mathcal{L}$, see (\ref{T-TOT}) and (\ref{T-split}). Shifting the mass replaces $m\to m-\delta m$ in the Lagrangian term of $P^e_\mu$. This is equivalent to adding a $\phi^2$ counterterm to $P^e_\mu$, see (\ref{T-split}). The resulting shift in the expectation value is,
\be
	\langle P^e_\mu\rangle \to \langle P^e_\mu\rangle+\frac{(m-\delta m)^2-m^2}{2kp}k_\mu =  \langle P^e_\mu\rangle -\frac{m\delta m}{kp}k_\mu + \mathcal{O}(e^4)\;,
\ee
which, with the standard lightfront $\delta m$ in (\ref{dm}), precisely cancels the remaining divergence in (\ref{renorm-mig}). The extension to the finite terms is trivial. (That the mass-shift is the same as that required for renormalising $S$-matrix elements is a good sign.) The expectation value of the renormalised operator is, finally,
\be\begin{split}\label{ny-p}
	 \langle P^e_\mu\rangle_\text{renorm.} &= p_\mu + \mathcal{O}(e^4) \;,
\end{split}
\ee
which is the desired result\footnote{$\langle P^\gamma\rangle$ is also divergent, and to obtain a finite result $P^\gamma_\mu$ can mix with $P^e_\mu$~\cite{Sterman:1994ce,Brown:1992db}, giving $\langle P^\gamma\rangle  \to \langle P^\gamma\rangle-\frac{\alpha}{2\pi\epsilon}\langle P^e \rangle = 0 +\mathcal{O}(e^4)\;,$ the expected result. Another interpretation is that the split (\ref{T-split}) is corrected by quantum effects, and this requires redefining the momentum operators by mixing $P^e$ and $P^\gamma$: after transferring the $p_\mu$ divergence from $P^e$ to $P^\gamma$, and renormalising the electron mass, both operators yield finite expectation values.}. We have identified the renormalised momentum operator which yields the finite, physical electron momentum to order $e^2$. We now return to the theory with a background field.

\section{Radiation reaction at finite time}\label{SEKT:RR}
%
We now have everything required to calculate the expectation value (\ref{P-EXP-FRI}); we have the state (\ref{tillstaand-phi}), the momentum operator (\ref{P-split-3}), and we know how to renormalise. The above renormalisation yields finite results also in-background; no `new' UV divergences arise, see also~\cite{Becker:1976ne}. To order $e^2$, the expectation value comprises the following four terms
\be\label{p-op-expansion}\begin{split}
	\langle P^e_\mu \rangle &= \bra{\text{in}} P^{(0)}_\mu\ket{\text{in}} \\
	&+\bra{\text{in}}\int\limits^{x^\LCp}\!\ud y V_1(y) P^{(0)}_\mu \int\limits^{x^\LCp}\!\ud z V_1(z)\ket{\text{in}} - 2\text{Re}\, \bra{\text{in}} P^{(0)}_\mu \int\limits^{x^\LCp}\ud y\int\limits^{y}\ud z V_1(y) V_1(z)\ket{\text{in}}  \\
	&+2\text{Im}\, \bra{\text{in}}P^{(1)}_\mu\int\limits^{x^\LCp}\!\ud y^\LCp\ V_1(y^\LCp)\ket{\text{in}} \;,
\end{split}
\ee
which we write as, respectively,
\be
	\langle P^e_\mu \rangle= \langle \text{free}\rangle_\mu  + \langle \text{emission}\rangle_\mu +\langle \text{loop}\rangle_\mu+ \langle \text{operator}^{(1)}\rangle_\mu \;.
\ee
These names refer to the origin and physical meaning of each term, as we now describe. 

\subsubsection*{Free}
\begin{figure}[t]
\centering\includegraphics[width=0.25\textwidth]{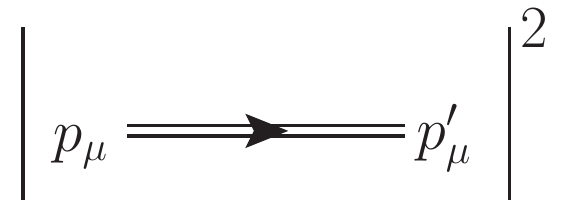}
\caption{\label{fig-noll} The Furry-Feynman diagram describing the zeroth order (Lorentz-force) contribution to the average electron momentum, $\langle\text{free}\rangle_\mu$. Double lines indicate the background-dressed propagator.  (For simplicity we draw, here and below, the diagrams of the covariant theory, rather than lightfront time-ordered diagrams.) }
\end{figure}

$\langle\text{Free}\rangle$ is the contribution from the `free' theory without QED interactions. It corresponds to scattering-without-emission, ${\mathrm e}^-\to {\mathrm e}^-$, in which the electron is accelerated by the background, but the electron does not emit.  It should therefore yield the Lorentz force component. From (\ref{tillstaand-noll}) and (\ref{P-split-3}), we find
\be
\begin{split}
	\langle \text{free}\rangle_\mu &= \bra{\text{in}} P^{(0)}_\mu \ket{\text{in}} = \int\!\ud{\sf p}\ |g({\sf p})|^2 \pi_\mu(\phi)  \to \pi_\mu(\phi) \;,
\end{split}
\ee
where in the final step we have taken the limit in which the wavepacket is strongly peaked; this is indeed the Lorentz force result.  See Fig.~\ref{fig-noll} for the associated diagram.

%
\subsubsection*{Loop and emission}
\begin{figure}[t]
\centering\includegraphics[width=0.5\textwidth]{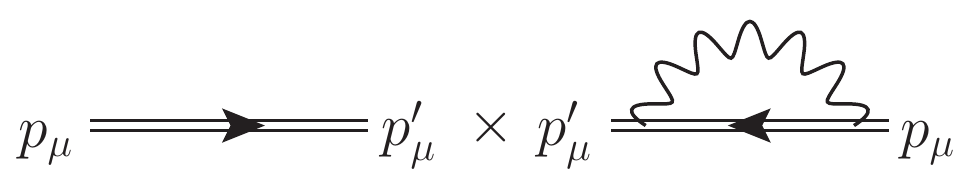}\centering\includegraphics[width=0.5\textwidth]{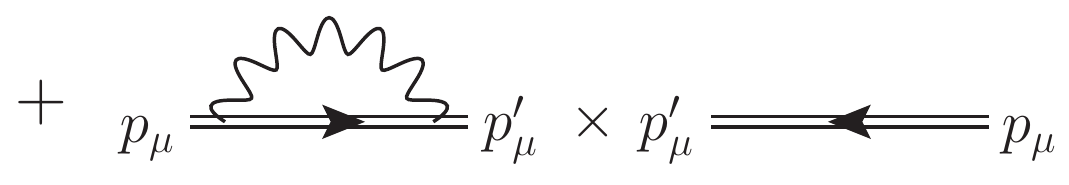} \\
\includegraphics[width=0.3\textwidth]{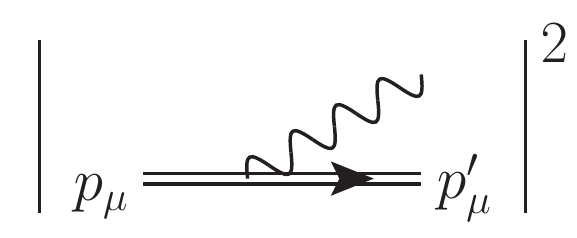}
\caption{\label{fig-ett} Furry-Feynman diagrams for two of the $\mathcal{O}(e^2)$ contributions to the average electron momentum, and to radiation reaction. In the first line, the cross term from scattering-without-emission, and in the second line, nonlinear Compton scattering, at tree level, mod squared.}
\end{figure}

%
$\langle\text{Loop}\rangle$ is the tree-level/one-loop cross term from scattering-without-emission, i.e.\ a contribution from the self energy of the electron.  $\langle\text{Emission}\rangle$ corresponds to single photon emission from the electron, which in a plane wave background is called nonlinear Compton scattering~\cite{Nikishov:1963,Nikishov:1964a,Harvey:2009ry,Boca:2009zz,Heinzl:2009nd,Seipt:2010ya,Mackenroth:2010jr,Boca:2012pz}.  See the diagrams in Fig.~\ref{fig-ett}. To calculate the emission and loop terms, we need the result, writing $\ud{\sf l}'\equiv \ud{\sf l}\,\theta(kp-kl)$,
\be\label{H-in}
	\int\limits^{x^\LCp}\! \ud y^\LCp V_1(y^\LCp) \ket{\text{in}} = e\!\int\limits^\phi\!\ud\phi_1\!\int\!\frac{\ud{\sf l}'\ud{\sf p}}{kp-kl}\ g({\sf p}) \pi_{\sf p}a^\dagger ({\sf l}) b^\dagger({\sf p}-{\sf l})\ket{0} 	e^{i\int\limits^{\phi_1}\frac{l\pi}{kp-kl}} \;.
\ee
We find that the two terms differ only in their vector structure, and a relative minus sign. (This is a direct consequence of the optical theorem.) $\langle\text{Loop}\rangle_\mu$ contains the Lorentz force (no recoil) result $\pi_\mu(\phi)$ while $\langle\text{emission} \rangle$ contains $p'_\mu$, the electron's momentum following photon emission in nonlinear Compton~\cite{Harvey:2009ry,Boca:2009zz,Heinzl:2009nd,Seipt:2010ya,Mackenroth:2010jr,Boca:2012pz,Ilderton:2013tb}:
\be
	p_\mu'(\phi) = \pi_\mu(\phi) - l_\mu+\frac{l\pi(\phi)}{kp-kl}k_\mu \;.
\ee
Reinstating $\hbar$, we find
\be\label{emission-loop}
\langle \text{loop + emission} \rangle_\mu =\frac{e^2}{\hbar^3kp}\int\!\ud{\sf l}'\ \big[\pi_\mu(\phi)-p'_\mu(\phi)\big] \int\limits^\phi\!\ud\phi_1\ud\phi_2\, \frac{\pi_2L\pi_1}{kp-kl} e^{\frac{i}{\hbar}\int\limits_{\phi_1}^{\phi_2}\frac{l\pi}{kp-kl}} \;.
\ee
The reason we group the loop and emission terms together is that while both are individually infra-red (IR) divergent~\cite{Dinu:2012tj}, their sum is IR finite~\cite{Ilderton:2013tb}. Ths is because, unlike individual $S$-matrix elements, our expectation value is an inclusive observable (no assumption is made about the `final' state)~\cite{Bloch,Kinoshita:1962ur,Yennie:1961ad}. The potentially IR-divergent term in $\langle\text{loop}\rangle$ is removed by $\langle\text{emission}\rangle$, due to the vector structure $\pi_\mu-p'_\mu$. See \cite{Gelis:2013oca} for related comments regarding the Schwinger mechanism.

The sum $\langle\text{loop + emission}\rangle$ does though contains a UV divergence. We can separate out the divergent term using integration-by-parts to rewrite the $\phi$-integrals as follows
\begin{align}
\label{I-INT}
\langle &\text{loop $+$ emission}\rangle_\mu = \\
\nonumber  &e^2\int\!\ud{\sf l}' \frac{kp-kl}{\hbar kp}\, \big[\pi(\phi)-p'(\phi)\big]_\mu \left|\frac{\pi(\phi)-\tfrac{kp}{kl}l}{l\pi(\phi)}-\int\limits^\phi\ud\phi_1\bigg(\frac{\pi_1-\tfrac{kp}{kl}l}{l\pi_1}\bigg)'\exp \frac{i}{\hbar}\int\limits_{\phi_1}^\phi\frac{l\pi}{kp-kl}\right|^2 \;.
\end{align}
The boundary terms cannot be dropped: they lead both to UV divergences which must be renormalised and to finite, physical terms which must be retained. They are also essential for distinguishing between different equations in the classical limit. (Boundary terms were also crucial for the comparison between LAD and QED in~\cite{Higuchi:2002qc}.)  Expanding the modulus we obtain three terms,
\begin{align}\label{fult}
\langle &\text{loop $+$ emission}\rangle_\mu = e^2\int\!\ud{\sf l}' \frac{kp-kl}{\hbar kp}\, \big[\pi(\phi)-p'(\phi)\big]_\mu \bigg(\frac{\pi(\phi)-\tfrac{kp}{kl}l}{l\pi(\phi)}\bigg)^2 \\
\label{kant}&- 2 \text{Re}\, e^2\int\!\ud{\sf l}' \frac{kp-kl}{\hbar kp}\, \big[\pi(\phi)-p'(\phi)\big]_\mu  \bigg(\frac{\pi(\phi)-\tfrac{kp}{kl}l}{l\pi(\phi)}\bigg)\int\limits^\phi\ud{\phi_1} \bigg(\frac{\pi_1-\tfrac{kp}{kl}l}{l\pi_1}\bigg)'e^{\frac{i}{\hbar}\int\limits_{\phi_1}^\phi\frac{l\pi}{kp-kl}} \\
\label{kropp}&+ e^2\int\!\ud{\sf l}' \frac{kp-kl}{\hbar kp}\, \big[\pi(\phi)-p'(\phi)\big]_\mu \int\limits^\phi\ud{\phi_1}\ud\phi_2 \bigg(\frac{\pi_1-\tfrac{kp}{kl}l}{l\pi_1}\bigg)'\bigg(\frac{\pi_2-\tfrac{kp}{kl}l}{l\pi_2}\bigg)' e^{\frac{i}{\hbar}\int\limits_{\phi_1}^{\phi_2} \frac{l\pi}{kp-kl}}
\end{align}
The first line, (\ref{fult}), comes from the `boundary-boundary' terms of the two integrals, and is UV divergent. We will renormalise below, but first we write down the final contribution to our expectation value.
\subsubsection*{Operator}
Finally, $\langle\text{operator}^{(1)}\rangle$, is the contribution from the interacting part of the electron momentum operator, $P^{(1)}_\mu$. This contribution vanishes in the asymptotic limit (see below) and does not have a standard Feynman diagram description. Again using (\ref{H-in}) we find
\be\begin{split}\label{operator00}
\langle\text{operator}^{(1)}\rangle_\mu = 2\text{Im}\ \frac{e^2}{kp\hbar^2}\int\!\frac{\ud{\sf l}'}{kp-kl}\int\limits^\phi\!\ud\phi_1\ \bigg[\frac{1}{2}(2kp-kl)L_{\mu\nu}\pi^\nu_1  - k_\mu \pi(\phi)L\pi_1\bigg]e^{-\frac{i}{\hbar}\int\limits_{\phi_1}^\phi\frac{l\pi}{kp-kl}}.
\end{split}
\ee
We write the vector structure in (\ref{operator00}) as $V_{\mu}(\phi,\phi_1)$ to compactify notation. Integrating by parts to isolate the boundary term as in (\ref{fult})--(\ref{kropp}) gives
\begin{align}
\label{operator0}\langle\text{operator}^{(1)}\rangle_\mu &= -2\frac{e^2}{kp\hbar}\int\!{\ud{\sf l}'}\ \frac{V_\mu(\phi,\phi)}{l\pi(\phi)}\\
\label{operator} &+ 2\text{Re}\ \frac{e^2}{kp\hbar}\int\!{\ud{\sf l}'}\int\limits^\phi\!\ud\phi_1\ e^{-\frac{i}{\hbar}\int\limits_{\phi_1}^\phi\frac{l\pi}{kp-kl}}\frac{\partial}{\partial\phi_1}\bigg( \frac{V_\mu(\phi,\phi_1)}{l\pi(\phi_1)}\bigg) \;.
\end{align}
Again, the boundary term (\ref{operator0}) is UV divergent. Collecting all our terms together and regularising, the momentum integrals in (\ref{fult}) and (\ref{operator0}) can be performed exactly, and one finds
\be\label{allting}
	\langle \text{loop+emission+operator}^{(1)}\rangle_\mu = - \frac{\alpha}{2\pi\epsilon}\pi_\mu(\phi) + \frac{\alpha}{2\pi\epsilon}  \frac{m^2}{kp}k_\mu +(\ref{kant}) + (\ref{kropp}) + (\ref{operator}) \;.
\ee
The two divergences have precisely the same form as those without background. The first is proportional to the `free'-theory momentum $\pi_\mu(\phi)$, the second is proportional to $k_\mu$ and depends on $k\pi=kp$.  The same multiplicative and mass renormalisations as above remove the two divergences in (\ref{allting}). The remaining three terms, (\ref{kant}) and (\ref{kropp}) and (\ref{operator}) are UV finite, and unaffected by the renormalisation. These three terms are lowest order quantum recoil effects.  See \cite{US3} for numerical investigations.
%
%
%
\subsection{The classical limit}
%
The simplest way to take the classical limit follows from noting that photon momentum has no classical analogue, whereas wavenumber does: we therefore write $l_\mu = \hbar k'_\mu$ with $k'_\mu$ the outgoing photon's wavenumber. $\hbar$ can then be taken to zero in, (\ref{kant}), (\ref{kropp}) and (\ref{operator}),  and the integrals performed to obtain the classical result.

To better understand the physics of this limit, though, it is helpful to first highlight the dependence of the quantum results on the longitudinal momentum component $kl=\hbar kk'$. Longitudinal momentum is conserved by the Lorentz force, but this symmetry is broken by photon emission~\cite{Harvey:2011dp}; this non-conservation is seen explicitly when one goes from the solution of Lorentz to the solution of e.g.\ LAD or LL, see above and~\cite{Exact}.  Define the ratio of outgoing momenta by $\hbar t := \hbar kk' / kp' $, which may be equivalently written
\be\label{t}
	\hbar t = \frac{\hbar kk'}{kp-\hbar kk'} = \frac{kl}{kp-kl}  \;.
\ee
We then change integration variables from $l_\LCm$ to $t$, and we scale out the longitudinal momentum from $l_\mu$ by defining $r_\mu =  \frac{kp}{kl} l_\mu$, which amounts to a second change of variables, $l_\perp \to r_\LCperp = (kp/kl) l_\LCperp$. We illustrate this using the transverse component of (\ref{kropp}), which may be equivalently written
\be\label{KVANT-SD2}
(\ref{kropp}) = \frac{e^2}{4\pi}\!\int\limits_0^{\infty}\!\frac{\ud t}{(1+\hbar t)^3}\!\int\!\frac{\ud^2r_\LCperp}{(2\pi)^2}\ r^\LCperp \int\! \ud \phi_1\ud\phi_2 \cos\bigg(t \! \int\limits_{\phi_1}^{\phi_2} \frac{r\pi}{kp}\bigg)  \bigg(\frac{\pi_1-r}{r\pi_1}\bigg)'\bigg(\frac{\pi_2-r}{r\pi_2}\bigg)'  \;,
\ee
in which the symmetry of the integrals allows us to replace $\exp$ with $\cos$. Factors of $\hbar$ are removed from the exponents and now occur only in the combination $1+\hbar t$. The classical limit therefore corresponds to setting $\hbar=0$ in these factors. From (\ref{t}), this means dropping quantum effects associated with high energy photon emission. In this limit, the $t$-integral in (\ref{KVANT-SD2}) can be performed and yields a delta function,
\be\label{t-integralen}
	\int\limits_0^\infty\!\ud t \cos\bigg( t\int\limits_{\phi_1}^{\phi_2} \frac{r\pi}{kp}\bigg) = \pi\frac{kp}{r\pi(\phi_2)}\delta(\phi_2-\phi_1) \;,
\ee
which says that the coherent lightfront time integrals in (\ref{KVANT-SD2}) become incoherent in the classical limit. This means that interference terms, i.e.\ quantum effects, drop out~\cite{Ilderton:2013tb,Dinu:2013hsd}. (For a review of decoherence and the classical limit, see~\cite{Zurek:2003zz}.) Performing one of the $\phi$-integrals leaves us with
\be
	\lim_{\hbar\to 0} (\ref{kropp})  =  \frac{e^2kp}{16\pi^2}\int\!\ud r_\LCperp^2\  r^\LCperp\int\limits^\phi\! \ud \varphi\ \frac{{\pi'}^2}{(r\pi)^3} + m^2\frac{(r\pi')^2}{(r\pi)^5} =  \frac{2}{3} \frac{e^2}{4\pi}\frac{kp}{m^4} \int\limits^\phi a^{\prime\, 2} \pi^\LCperp\;,
\ee
which we recognise from the classical theory. (To obtain the final expression use $\pi r=\frac{1}{2}((r-\pi)_\LCperp^2+m^2)$ and change variables to $u_\LCperp = (r-\pi)_\LCperp$. The $u_\LCperp$ integrals are elementary.) The extension of the above calculation to the remaining terms, (\ref{kant}) and (\ref{operator}), is direct. We find the following classical limits for our various terms:
\be\begin{split}
\lim_{\hbar\to0} \langle\text{free}\rangle_\mu &=\pi_\mu(\phi) \\
\lim_{\hbar\to0}\ \langle\text{loop + emission}\rangle_\mu  &=\frac{1}{4}\Delta \pi'_\mu(\phi) + \frac{\Delta}{m^2}\int\limits^\phi a^{\prime\, 2} \bigg(\pi_\mu-\frac{\pi\pi(\phi)}{kp}k_\mu\bigg) \;,
\\
\lim_{\hbar\to0}\langle\text{operator}^{(1)}\rangle_\mu &=\frac{3}{4}\Delta \pi'_\mu(\phi) \;.
\end{split}
\ee
Summing these three contributions, one finds the final result
\be\label{RESULTAT-P}
	\lim_{\hbar\to 0}\ \langle P^e_\mu\rangle=q_\mu(\phi) \quad\text{ as in LAD, LL, EFO} \,.
\ee
We have obtained classical RR directly from the $\hbar\to 0$ limit of QED. The classical momentum of a radiating particle agrees with that predicted by the LAD, LL and EFO equations. We have made no approximation up to this point except that we work to order $e^2$ in the coupling; the extension to higher orders is discussed at the end of this paper.  We close this section with some comments on our result.

1) The $\hbar\to 0$ limit depends on which quantities (e.g.\ coupling, momenta) are chosen to scale with $\hbar$~\cite{Holstein,Brodsky:2010zk}. Since we wish to compare with the classical theory, where one has $e^2$ but no $\hbar$, we used the `standard' parameterisation, so for example $e^2$ is independent of $\hbar$, compare~\cite{Brodsky:2010zk}. It is not a problem that $\alpha = e^2/(4\pi\hbar)$ then diverges in the classical limit, since photon emission always brings with it a factor $\hbar$. It is the cancellation of $\hbar$ between the photon momentum and the coupling which leaves a nonzero classical result.

2)  In the context of deriving classical RR from QFT, it was noted in~\cite{Higuchi:2004pr} that loop terms could be relevant in the classical limit, see also~\cite{Holstein}.  We saw above that $\langle\text{loop}\rangle_\mu$ removes the $\pi_\mu$ term from $p'_\mu$. Had this term not been removed, it would have caused not only an IR divergence, but would have been proportional to $1/\hbar$, which blows up in the classical limit~\cite{Ilderton:2013tb}. (The UV divergences we encountered were also $1/\hbar$ terms).

3) The $\pi_\mu$ divergence we encountered is analogous to the classical divergence found in the derivation of LAD~\cite{Coleman}, which is also proportional to the momentum, or four-velocity. While the classical divergence is removed by mass renormalisation, it is removed by operator renormalisation in the quantum theory. The $k_\mu$ divergence and mass renormalisation in the quantum calculation are not seen classically, but are typical of lightfront quantisation.

4) For infinite lightfront times asymptotic switching kills $\langle\text{operator}^{(1)}\rangle$ and the boundary terms, including the divergences,  when integrating by parts. See \cite{Boca:2009zz,Heinzl:2009nd,Mackenroth:2010jr} for the explicit calculation with exponential damping factors in nonlinear Compton scattering. (This regularisation is gauge invariant~\cite{Ilderton:2010wr}.) The remaining term (\ref{kropp}), from the sum of nonlinear Compton scattering and scattering-without-emission in Fig.~\ref{fig-ett}, becomes
\be\label{KVANT-SD}
	\langle P^e_\mu\rangle\bigg|_{\phi=\infty} =\hat\pi_\mu + \displaystyle\frac{e^2}{4\pi\hbar}\int\limits_0^{p_\LCm}\!\frac{\ud l_\LCm}{l_\LCm}\!\int\!\frac{\ud^2l_\LCperp}{(2\pi)^2} \frac{kp'}{kp}  \big(\hat{\pi}_\mu-p'_\mu\big) \! \int\!\ud\phi_1\ud\phi_2 e^{\frac{i}{\hbar}\int\limits_{\phi_1}^{\phi_2}\frac{l\pi}{kp'}}\partial_2\partial_1 \bigg(\frac{\pi_2\pi_1}{l\pi_2 l\pi_1} \bigg) \;,
\ee
which agrees with the QED $S$-matrix result of~\cite{Ilderton:2013tb} if one neglects spin effects. (Set $g=1/2$ in \cite{Ilderton:2013tb}  to go from spinor to scalar QED.)  We remark that if one knew only this asymptotic result, and assumed it to be valid at finite time just by changing the $\phi$-integral limits, one would miss $\langle\text{operator}^{(1)}\rangle$ and the boundary term (\ref{kant}). These are precisely the terms which become $\Delta\pi'$ when $\hbar\to 0$ and distinguish between the different classical equations\footnote{This seems to be the reason~\cite{Sokolov:2010jx} claims that S follows from QED.}.

Recalling the classical discussion (\ref{asympt})-(\ref{uppfattning1}), we consider the component $k\langle P^e\rangle$. From (\ref{KVANT-SD}) we find
\begin{equation}
	\frac{k\langle P^e\rangle}{kp}\bigg|_{\phi=\infty} = 1 + \alpha\int\limits_0^{p_\LCm}\!\frac{\ud l_\LCm}{l_\LCm}\!\int\!\frac{\ud^2l_\LCperp}{(2\pi)^2} \bigg[\frac{kl\, kp'}{kp^2}\bigg]  \! \int\!\ud\phi_1\ud\phi_2 e^{\frac{i}{\hbar}\int\limits_{\phi_1}^{\phi_2}\frac{l\pi}{kp'}}\partial_2\partial_1 \bigg(\frac{\pi_2\pi_1}{l\pi_2 l\pi_1} \bigg) \;,
\end{equation}
where the term in square brackets shows that the integrand depends on the product of the outgoing longitudinal momenta, in ratio to the incoming momentum squared. Hence it seems natural to use $k\langle P^e\rangle/kp$ to estimate when higher order effects become important, phenomenologically, in analogy to using $kq/kp$ in the classical theory. The relative importance of classical and quantum effects, and when these are important compared to higher order effects in $\alpha$, will be addressed in detail in~\cite{US3}.
\subsection{Current and position}\label{SEKT:X}
The above calculation can be extended to the current. The classical current for a particle with orbit $x^\mu_\text{cl}$ is 
\be
	J_\text{cl}^\mu(x)=\int\!\ud\tau\ \delta^4(x-x_\text{cl}(\tau))\frac{\ud x^\mu_\text{cl}}{\ud \tau} \;.
\ee
Changing variables from $\tau$ to $\phi$ and integrating over the spatial coordinates gives
\be\label{current}
	\frac{\ud x^\mu_\text{cl}}{\ud\phi}=\frac{1}{2 k_\LCp}\int\! \ud{\sf x}\ J_\text{cl}^\mu({\sf x},\phi) \;,
\ee
the right hand side of which can be compared to the classical limit of the current operator's expectation value. Here we need the full current (compare with (\ref{current}))
\be
	J_\mu=\Phi^\dagger (i\overset{\leftrightarrow}{\partial}_\mu-2a_\mu-2eA_\mu)\Phi \;,
\ee
and $A_\LCp$ in this equation is as in (\ref{constraint}).  The calculation proceeds as above, and there is a $\phi$-independent divergence which is removed by mass renormalisation. Rather than go through this, we present a calculation which yields the same result, but is based on the position operator. 
Localisation~\cite{RS} and position~\cite{NW} in relativistic QFT are difficult subjects with a long history, see~\cite{Duncan}. While it is safe to say that position is not as natural a variable as momentum, it is still interesting to consider the position operator, and we will see that our calculation yields results consistent with (\ref{RESULTAT-P}).

As above, we will calculate the position operator's expectation value and take $\hbar\to 0$. The idea of~\cite{Higuchi:2002qc} is that, at least to lowest order and in instant-form quantisation, one can use the charge density $J_0$ to define a position operator by
\be\label{Higuchi}
	\hat{{\bf x}} =\int\!\ud^3 x\ {\bf x}\, J_0 \;.
\ee
For one-electron states with wavepackets $g_2$ and $g_1$ it is straightforward to show, again in instant-form quantisation, that this operator obeys (setting $x^0=0$ for simplicity)
\be
	\bra{g_2}\hat{\bf x}^j\ket{g_1} = \int\!\frac{\ud^3{\bf p}}{(2\pi)^32E_{\bf p}} g^*_2({\bf p})\bigg(-i\frac{\partial}{\partial {\bf p}_j}-i\frac{{\bf p}^j}{2E_{\bf p}^2}\bigg)g_1({\bf p}) = (g_2,{\bf Q}_\text{NW}g_1) \;,
\ee
in which we recognise the Newton-Wigner position operator ${\bf Q}_\text{NW}$~\cite{NW}. The position operator (\ref{Higuchi}) was applied to RR, for a different background than ours, in the papers~\cite{Higuchi:2002qc,Higuchi:2004pr,Higuchi:2005an,Higuchi:2009bh}. Here we extend those calculations to our background, and to lightfront quantisation, in which the charge density is $J_\LCm=j_\LCm$. The obvious extension of the position operator (\ref{Higuchi}) to lightfront coordinates would then seem to be
\be\label{X-uttryck}
	{{\sf X}}(\phi) = \int\!\ud{\sf x}\ {\sf x}\, j_\LCm(\phi) \;.
\ee
We are immediately confronted by the results of~\cite[\S 2E]{Leutwyler:1977vy}. That paper extends the definition of the Newton-Wigner position operator to the front form, by writing down constraints on the algebra of a candidate position operator with the stability group of the null plane $x^\LCp=0$. It is then concluded that no self-adjoint longitudinal position operator exists which satisfies these constraints. Interesting questions for future study are then to what extent does ${\sf X}$ satisfy the front-form Newton-Wigner conditions of~\cite{Leutwyler:1977vy}, and what is the physics of any difference? Here we note that the covariant measure in $\ud{\sf p}$, and hence in ${\sf X}$, is precisely that which yields the ``unique notion of locality'' identified in~\cite[\S 3]{Leutwyler:1977vy}, at least for elementary systems. Given this and that (\ref{X-uttryck}) seems like a natural place to start,    we will proceed to analyse $\langle{\sf X}\rangle$ and comment further on the longitudinal component below.

We begin with a check on the definition (\ref{X-uttryck}) by calculating the simplest component, obtained by replacing ${\sf x}$ in (\ref{X-uttryck}) with $x^\LCp$. In other words, we wish to check that our position operator correctly generalises to the lightfront time component. In this case, the contributing terms of the `position' operator (\ref{X-uttryck}) are
\be
	x^\LCp \int\!\ud{\sf x}\ j_\LCm(x^\LCp) = x^\LCp \int\!\ud{\sf p}\ b^\dagger({\sf p})b({\sf p}) \;,
\ee
which counts the number of electrons in the system and then multiplies by $x^\LCp$. Since we have only one electron, and net fermion number is conserved, the expectation value becomes
\begin{align}
	\langle {X}^\LCp\rangle = x^\LCp \bracket{\psi;x^\LCp}{\psi;x^\LCp} = x^\LCp\bracket{\text{in}}{\text{in}} =x^\LCp \;,
\end{align}
as it should be. With this checked, we return to the transverse and longitudinal position operators (\ref{X-uttryck}). The expectation value of interest is
\begin{align}
	\nonumber \langle {\sf X} \rangle &= \bra{\text{in}}{\sf X}\ket{\text{in}}+\bra{\text{in}}\int\limits^{x^\LCp}\!\ud y V_1(y)\, {\sf X} \!\int\limits^{x^\LCp}\!\ud z V_1(z)\ket{\text{in}} - 2\mathrm{Re}\bra{\text{in}}{\sf X} \int\limits^{x^\LCp}\ud y\!\int\limits^{y}\!\ud z V_1(y) V_1(z)\ket{\text{in}}  \\
\label{x-op-exp}	&=: \langle\text{free}\rangle  + \langle \text{emission}\rangle  + \langle \text{loop}\rangle \;.
\end{align}
In $A^\LCp=0$ gauge there is no $\langle\text{operator}^{(1)}\rangle$ contribution. As above, $\langle \text{free}\rangle$ is the `free' theory contribution. To calculate it we first evaluate the expectation value of the current in the initial state,
\be
	\bra{\text{in}} j_\LCm(x^\LCp) \ket{\text{in}} = \int\!\ud{\sf p}\ud{\sf q} \ g^*({\sf q})g({\sf p})(p+q)_\LCm \varphi^*_{\sf q}(x^\LCp) \varphi_{\sf p}(x^\LCp) \;,
\ee
and then perform the ${\sf x}$-integrals in (\ref{X-uttryck}), arriving at
\be\label{XX}
	\langle \text{free}\rangle=\bra{\text{in}}{\sf X}\ket{\text{in}}=\int\!\ud{\sf p}\ |g({\sf p})|^2\int\limits^\phi\frac{\sfpi}{kp}-ig({\sf p})^*\frac{\partial}{\partial{\sf p}} g({\sf p}) \;,
\ee
in which $\raisebox{3pt}{\sfpi}=\{\pi^\LCperp,\pi^\LCm\}$. For both ${\sf X}^\LCperp$ and the longitudinal component ${\sf X}^\LCm$, we have the expected particle momentum in the first term of (\ref{XX}), while the second term is time-independent and is the only surviving term at the initial time. We therefore associate it with the initial position. In the limit that $g$ becomes strongly peaked we have
\be
	\langle \text{free}\rangle = {\sf x}_\text{init} + \int\limits_0^\phi\frac{\sfpi}{kp} \implies	\frac{\ud}{\ud \phi}\langle {\sf X}\rangle = \frac{1}{kp}{\rotatebox[origin=c]{-90}{$\mathbf\vDash$}}
(\phi)+\mathcal{O}(e^2)\;,
\ee
which are the Lorentz force results for the position and momentum of a particle in a plane wave, as a function of lightfront time $\phi$. (As for the momentum calculation, the `free' result is independent of $\hbar$; the expectation value is equal to  the classical result.) 

$\langle \text{Loop}\rangle$ is again a finite time one-loop self-energy contribution. Using (\ref{H-in}) once more,
\be
	 \langle \text{loop}\rangle = 2\text{Re}\ \frac{e^2}{\hbar^3kp}\int\limits^{\phi}\!\ud \phi_2 \int\limits^{\phi_2}\ud\phi_1\int\!\ud{\sf l}' \frac{\pi_2L\pi_1}{(kp-kl)}\bigg( {\sf x}_\text{init}+\int\limits^{\phi}_0\frac{\sfpi}{kp} \bigg)e^{\frac{i}{\hbar}\int\limits_{\phi_1}^{\phi_2} \frac{l\pi}{kp-kl}}\;.
\ee
The final term, $\langle \text{emission}\rangle$, comes from nonlinear Compton scattering at tree level. In the sequel, we present the detailed calculation of the transverse co-ordinates, setting ${\sf x}_\text{init}=0$ in order to to simplify the presentation. Using (\ref{H-in}) again, one arrives after a straightforward calculation at
\begin{align}
\label{ta-limit-nu}	\langle {\sf X}\rangle^\LCperp = \frac{e^2}{2\hbar^3kp}\int\limits^\phi\! \ud\phi_2\ud\phi_1 \int\!\ud{\sf l}'&\frac{1}{kp-kl} \bigg[\hbar(a_1-a_2)^\LCperp \sin\bigg(\frac{1}{\hbar}\int\limits_1^2\frac{l\pi}{kp-kl}\bigg) \\
\nonumber &+\frac{\pi_2L\pi_1}{kp-kl} \cos\bigg(\frac{1}{\hbar}\int\limits_1^2\frac{l.\pi}{kp-kl}\bigg)\bigg[\int\limits_1^\phi+\int\limits_2^\phi\bigg]\bigg(l-\frac{kl}{kp}\pi\bigg)^\LCperp\bigg]
\end{align}
In this case there are no contributions from boundary terms, and no UV-divergences. (See \cite[\S 11]{Sterman:1994ce} and \cite{Collins:2005nj} for renormalisation of the current in QED). The classical limit is obtained as above.  Taking a $\phi$ derivative to simplify the presentation and compare directly with the classical calculations, the first and second lines of (\ref{ta-limit-nu}) contribute, respectively,
\be
	\frac{3}{4}\Delta {\pi'}^\LCperp(\phi) \;, \qquad \frac{1}{4}\Delta {\pi'}^\LCperp(\phi) + \frac{\Delta}{m^2}\int\limits^\phi a'^2\big(\pi-\pi(\phi)\big)^\LCperp \;.
\ee
Summing these two, we find the transverse position given in (\ref{LAD-X}) by a subset of the classical equations. In fact, this is enough to distinguish between the different classical equations, but we can nevertheless repeat the calculation for the longitudinal direction. This calculation is technically more involved due to the $p_\LCm$ derivatives acting on the many factors of $kp\sim p_\LCm$ throughout our expressions, but no other difficulties arise. One finds the final result
\be\label{RESULTAT-X}
	\lim_{\hbar \to 0} \frac{\ud}{\ud\phi}\langle {X}^\mu\rangle= \frac{\ud}{\ud\phi}x^\mu(\phi)\quad \text{ as in LAD, LL, EFO, S} \;.
\ee
Combining the two results (\ref{RESULTAT-X}) and (\ref{RESULTAT-P}), we see that, of the classical equations considered above, only three,  LAD, LL and EFO are consistent with QED to this order. Further $q=m\dot{x}$ even with recoil effects.

\subsection{Higher orders}
\begin{figure}[t!]
\centering\includegraphics[width=0.6\textwidth]{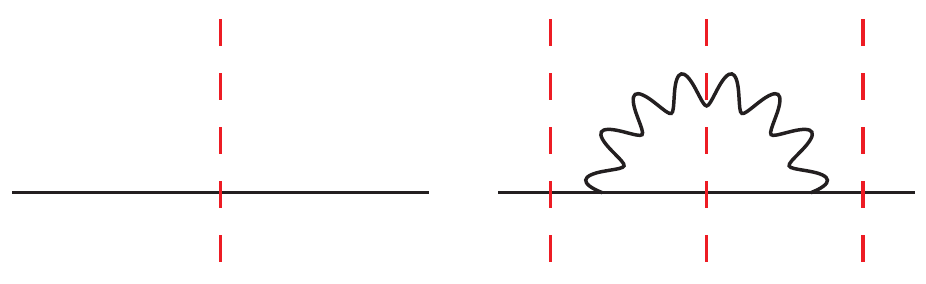}\caption{\label{HO-1} All possible cuts of these diagrams generate all possible terms to be calculated for $\langle kP^e\rangle$ to order~$e^2$: dashed vertical lines correspond to a cut and the insertion of $kP^e=kP^{(0)}$.}
\end{figure}
To distinguish between those equations which we have found to be consistent with QED, one needs to consider order $e^4$ effects at finite time. (All considered equations agree on the post-pulse, {\it asymptotic} momentum to order $e^4$, which can be obtained from the order $e^2$ result (\ref{LAD-P}) and Larmor's formula.) One option is to calculate $k\langle P^e\rangle$ as a function of lightfront time and compare it to the classical result 
\be
	\frac{kq(\phi)}{kp}= 1+\frac{\Delta}{m^2}\int\limits^\phi a'^2+\frac{\Delta^2}{m^4}\bigg(\int\limits^\phi a'^2\bigg)^2+C\, \frac{\Delta^2}{m^2}a'^2 \;,
\ee
where $C_\text{LAD}=2$, $C_\text{LL}=0$ and $C_\text{EFO}=1/2$. 

\begin{figure}[t!]
\centering\includegraphics[width=0.8\textwidth]{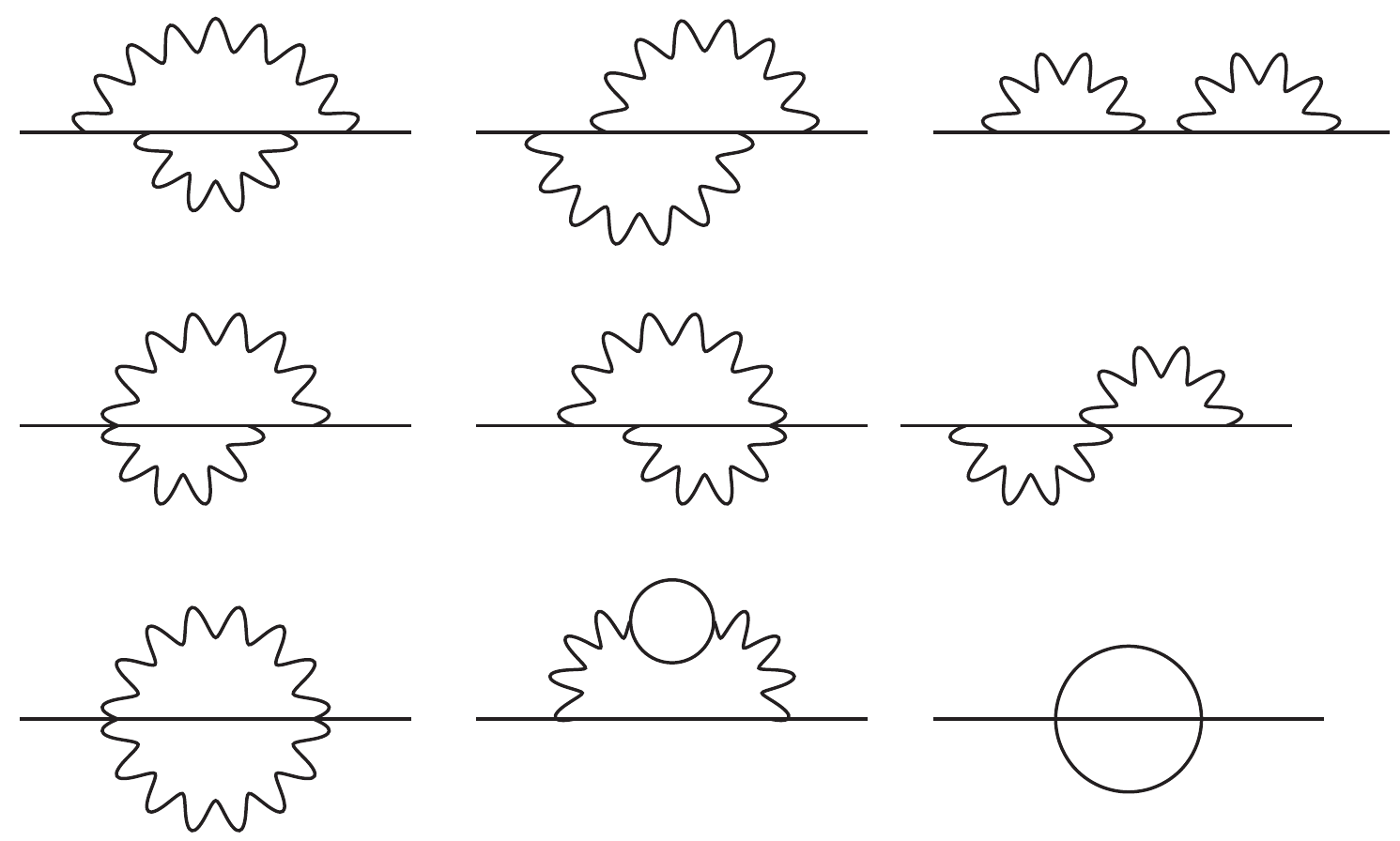}\caption{\label{HO-2} Order $e^4$ contributions to RR come from all possible cuts of these diagrams. (The required counterterms are not shown.) Vacuum polarisation and pair production effects are confined to all possible cuts of the second and third diagrams in the third line, the latter containing the four-scalar interaction.}
\end{figure}

We expect the next order calculation to be more difficult than that presented here, though. There are a few reasons for this. Consider $k\langle P^e\rangle=k\langle P^{(0)}\rangle$. (This component is simplest because the operator contribution drops out.) The two diagrams in Fig.~\ref{HO-1} represents all terms in this expectation value, to order $e^2$. A vertex is the three-point vertex in (\ref{H-ALL}) (including the instantaneous photon interaction). One obtains the terms to be calculated, including lightfront time-ordering, from all possible cuts of the diagrams (red dashed lines) with the operator $kP^{(0)}$ inserted at the cut.  The part of the diagram to the left (right) of the cut then belongs to the bra (ket). This notation serves to reflect the unity of different contributions; in the second diagram, the loops given by the left and right cuts must be retained and treated together with the emission diagram, obtained from the central cut, in order to remove IR divergences. There are many more terms to calculate at order $e^4$, obtained from all cuts of the diagrams in Fig.~\ref{HO-2}, with $kP^{(0)}$ inserted at the cut.

Most of the diagrams in Fig.~\ref{HO-2} describe multi-photon emissions, along with loop corrections to those emissions. The loops cannot be dropped from the outset. The second diagram in the third line contains effects from pair production and vacuum birefringence. Both are quantum processes, and so one expects that they should drop out as $\hbar\to 0$, or that there is a parameter regime in which they are much less important than those in the photon emission diagrams~\cite{DiPiazza:2010mv,Neitz}. If one is only interested in the classical limit, it should be sufficient to drop the pair-birefringence diagram (and the final diagram in Fig.~\ref{HO-2} which describes pair production from the instantaneous-fermion vertex). We believe that $1/\hbar$ terms cancel amongst themselves within the diagrams shown, i.e.\ within the {\it groups} of terms obtained from cutting the diagrams.  At higher orders, one might also expect that UV--divergent terms and order $\epsilon$ terms can combine, through $1/\epsilon*\epsilon=1$ cancellations, to give finite terms.

One can also compare QED results with the transverse components of the classical orbit, $\ud x^\LCperp/\ud\phi$. We believe it would be simplest to do so using the current operator, as in (\ref{current}) because in this case there are no divergences at order $e^2$ and the calculation is technically simpler than when using ${\sf X}^\LCperp$ (as there are no momentum derivatives to calculate).

In this paper we have focussed on the momentum of the electron. To see RR in the photon spectrum one also needs to consider at least an order $e^4$ calculation (requiring emission of two photons~\cite{Seipt:2012tn}, or more~\cite{DiPiazza:2010mv,Neitz}), since the photon spectrum contains no RR effects at order $e^2$~\cite{Ilderton:2013tb}. This is why no RR effects were seen in previous calculations of the photon momentum from nonlinear Compton scattering~\cite{Harvey:2009ry,Boca:2009zz,Heinzl:2009nd,Seipt:2010ya,Mackenroth:2010jr,Boca:2012pz}.

\section{Conclusions}\label{SEKT:SAMMANFATTNING}
We have derived dynamical quantum and classical radiation reaction from scalar QED. Calculating the expectation value of the momentum and position operators, we saw that radiation reaction appeared at order $\alpha$, due to photon emission and self-energy effects. The $\hbar\to 0$ limit of our expectation values gave us the orbit of a classical, radiating particle. Our results imply that, of the classical equations in Sect.~\ref{klass-sekt}, only LAD, LL and EFO can be consistent with QED. This is consistent with known first-order relations between these three equations~\cite{LL-bok,Spohn:1999uf,Hammond}, and with previous results on the derivation of RR from QED~\cite{Krivitsky:1991vt,Higuchi:2002qc}.

Our calculation used the Hamiltonian formalism, the Furry picture, and a plane wave background. This choice made lightfront field theory the natural  framework to adopt. Canonical quantisation of a particle in a plane wave is simple on the lightfront~\cite{Neville:1971uc}, see also~\cite{Ilderton:2012qe}, but it is only in recent years that progress has been made on instant-form quantisation of the same theory~\cite{BOCA,Lavelle:2013wx}. (Recall that even the Lorentz orbit of a particle in a plane wave has no closed-form parameterisation in instant-form time~$x^0$.)

Since expectation values are inclusive observables, all our expressions were infra-red finite. We found two ultra-violet divergences. The first was analogous to that found in the classical derivation of LAD, being proportional to the particle's momentum, and was removed by a multiplicative renormalisation. The second was particular to lightfront quantisation and was removed by the usual mass counterterm of lightfront QED. Here we again saw the suitability of lightfront quantisation for our problem; transverse dim reg~\cite{Casher:1976ae} can be used without affecting the structure of the chosen background field. 

In complete analogy to the usual coupling expansion of QED, we worked to a certain order in $\alpha$. We found that lowest order (classical) RR effects came from diagrams of lowest order in $\alpha$, meaning that we worked in the regime in which RR effects are small. We note that, aside from a few exact solutions~\cite{Exact}, this same expansion is almost the only approach available for treating RR in strong-field QED~\cite{DiPiazza:2011tq}. Calculations in the RR-dominated regime use the same expansion, but carried out to higher orders, see e.g.~\cite{DiPiazza:2010mv}. A fully quantum, non-perturbative approach is sadly lacking\footnote{Note that the majority of current numerical implementations of strong field QED are based not on a nonperturbative discretisation of QED, as would be the case in lattice gauge theory, but on the addition of the perturbative nonlinear Compton (and pair production) cross-sections to classical PIC codes.}. However, to identify the quantum origins of RR, or to rule out classical equations, working to low orders in $\alpha$ is sufficient. In fact, our methods can, already at order $e^4$, distinguish between LAD, LL and EFO. We expect the calculation to be significantly harder than that presented here due to the multitude of terms to be calculated, and new divergent structures which can appear.

\acknowledgments
We thank Chris Harvey, Tom Heinzl and Mattias Marklund for useful discussions. The authors are supported by the Swedish Research Council, contract 2011-4221.

\appendix

\section{Another classical equation}\label{S-APP}
The approach of \cite{SOK-JETP} (S) is to begin with two coupled equations
\be\label{Sok-CE}
	\dot{q}=mf\dot{x}-\dot{q}_r \;, \qquad \dot{x}=\frac{1}{m}q+\dot{x}_r \,.
\ee
in which $q^2=m^2$ but $\dot{x}^2\neq 1$. So momentum, which is on mass-shell, is not proportional to velocity, which is off mass-shell. With the particular choice~\cite{SOK-JETP}
\be
	\dot{q}_r=-\frac{2}{3}\frac{e^2}{4\pi}\frac{1}{m^3}(fq)^2q \;, \qquad \dot{x}_r=\frac{2}{3}\frac{e^2}{4\pi}\frac{1}{m^2} fq \;,
\ee
one finds
\be
	\dot{q}=fq+\frac{2}{3}\frac{e^2}{4\pi}\left(\frac{1}{m}ffq+\frac{1}{m^3}(fq)^2q\right) \;,
\ee
which looks like LL, but without the first term, see Table~\ref{R-tabel}. Thus, Sokolov's equations have the form (\ref{eom}) for the momentum, but not for the velocity.

\end{document}